\documentclass[reprint,amsmath,amssymb,aps,pra]{revtex4-1}
\usepackage{graphicx}
\usepackage{dcolumn}
\usepackage{bm}

\usepackage{amsmath} 
\usepackage{color}
\usepackage[colorlinks=true, linkcolor=blue,urlcolor=blue,anchorcolor=blue,citecolor=blue,bookmarksnumbered]{hyperref}

\begin{document}

\title{{Enhanced phonon blockade in a weakly-coupled hybrid system via mechanical parametric amplification}}

\author{Yan Wang$^1$}
\author{Jin-Lei Wu$^1$}
\author{Jin-Xuan Han$^1$}
\author{Yan Xia$^5$}
\author{Yong-Yuan Jiang$^{1,2,3,4}$}
\author{Jie Song$^{1,2,3,4}$} \email{E-mail: jsong@hit.edu.cn}
\affiliation{$^1$ School of Physics, Harbin Institute of Technology, Harbin, 150001, China\\
$^2$ Collaborative Innovation Center of Extreme Optics, Shanxi University, Taiyuan, 030006, China\\
$^3$ Key Laboratory of Micro-Nano Optoelectronic Information System, Ministry of Industry and Information Technology, Harbin, 150001, China\\
$^4$ Key Laboratory of Micro-Optics and Photonic Technology of Heilongjiang Province, Harbin Institute of Technology, Harbin, 150001, China\\
$^5$ Department of Physics, Fuzhou University, Fuzhou, 350002, China}

\date{\today}

\begin{abstract}
We propose how to achieve strong phonon blockade (PB) in a hybrid spin-mechanical system in the weak-coupling regime. We demonstrate the implementation of magnetically-induced two-phonon interactions between a mechanical cantilever resonator and an embedded nitrogen-vacancy (NV) center, which, combined with parametric amplification of the mechanical motion, produces significantly enlarged anharmonicity in the eigenenergy spectrum. In the weak-driving regime, we show that strong PB appears in the hybrid system along with a large mean phonon number, even in the presence of strong mechanical dissipation. We also show flexible tunability of phonon statistics by controlling the strength of mechanical parametric amplification. Our work opens up prospects for the implementation of an efficient single-phonon source, with potential applications in quantum phononics and phononic quantum networks.

\end{abstract}
\maketitle
\section{Introduction}
Over the past decade the study of hybrid quantum systems (HQSs) has attracted great interests in exploring new phenomena and developing novel quantum technologies \cite{RevModPhys.85.623}. By combining different physical components with
complementary functionalities, HQSs could provide multitasking capabilities that the
individual components cannot offer \cite{Kurizki3866}. With the recent progress in micro- and nanomechanical fabricating technologies, mechanical oscillators, such as suspended membranes or tiny cantilevers, have found diverse applications in a wide variety of disciplines \cite{mamin2007nuclear,doi:10.1021/nl902350b,Poggio_2010,chaste2012nanomechanical,C2NR31102J,doi:10.1063/1.3673910}. In particular, these mechanical elements can serve as an essential component of HQSs when being interfaced with some other quantum systems, such as superconducting qubits  \cite{lahaye2009nanomechanical,o2010quantum} or photons \cite{PhysRevLett.107.133601,verhagen2012quantum}. Another prominent example is the hybrid spin-mechanical systems \cite{PhysRevB.79.041302,rabl2010quantum}, which fully take advantages of the outstanding coherence property of solid-state spins and the high quality factors of nanomechanical oscillators \cite{DOHERTY20131,POOT2012273,doi:10.1063/5.0024001}, thus being widely studied for both fundamental and practical applications in quantum information science \cite{arcizet2011single,kolkowitz2012coherent,PhysRevLett.110.156402,PhysRevLett.111.227602,PhysRevLett.113.020503,PhysRevApplied.4.044003,PhysRevLett.116.143602,PhysRevLett.117.015502,PhysRevApplied.5.034010,PhysRevB.94.214115,macquarrie2017cooling,PhysRevA.94.053836,cai2017second,PhysRevLett.121.123604,PhysRevLett.120.213603,PhysRevA.98.052346,PhysRevLett.125.153602,PhysRevA.101.042313,PhysRevResearch.2.013121}. As an important initial step, the cooling of mechanical oscillators close to their quantum ground state has been achieved in different experimental settings \cite{o2010quantum,teufel2011sideband,chan2011laser,clark2017sideband}.

The study of hybrid spin-mechanical setups also facilitates in-depth exploration of the quantum nature in mechanical systems, allowing ones to observe rich mechanical quantum effects. A typical mechanical quantum effect is phonon blockade (PB), which is an analog of { the} photon blockade in the cavity QED platform \cite{PhysRevLett.79.1467,birnbaum2005photon,PhysRevLett.104.183601,PhysRevLett.107.063601,manjavacas2012plasmon,PhysRevLett.109.193602,PhysRevA.88.023853,PhysRevA.87.023809,PhysRevA.89.043818,PhysRevLett.118.223605,PhysRevLett.118.133604,PhysRevLett.121.043601,PhysRevA.101.013826,Wang_2020}, and is typically associated with quantum mechanical nonlinearity. A strong mechanical nonlinearity could enable the emergence of PB, i.e., blockade of the excitation of the subsequent phonons by resonantly absorbing the first one \cite{PhysRevA.82.032101,PhysRevB.84.054503,PhysRevA.93.013808}. The research on PB has progressed enormously in the last few years since it paves a crucial step for implementation of quantum control at the single-phonon level. Recent schemes have predicted that PB could be induced in systems with the mechanical resonator coupled to a qubit \cite{PhysRevLett.110.193602,PhysRevA.93.063861,PhysRevA.94.063853}, in systems with quadratic optomechanical interactions \cite{PhysRevA.96.013861,PhysRevA.95.053844,PhysRevA.98.023819,PhysRevA.99.013804}, or in a system with magnetically induced two-phonon nonlinear coupling \cite{PhysRevA.100.063840}. However, for these existing schemes moving into a strong nonlinear regime of the mechanical resonators requires either large quadratic optomechanical coupling rate or large qubit-resonator coupling rate, which are still hard to realize in most real scenarios. Also, { the mean phonon number in the mechanical resonators is generally small due to the weak mechanical drive ensuring the generation of strong PB}, debasing greatly the efficiency of single-phonon emission as a single-phonon source. Therefore, in terms of improving the performance of PB, seeking for efficient approach that can be free from the restriction of strong coupling and yield simultaneously a large mean phonon number is highly desirable. In additon, in contrast to the conventional PB (relying on anharmonic eigenenergy spectrum) \cite{PhysRevA.82.032101,PhysRevB.84.054503,PhysRevA.93.013808,PhysRevLett.110.193602,PhysRevA.93.063861,PhysRevA.94.063853,PhysRevA.96.013861,PhysRevA.95.053844,PhysRevA.98.023819,PhysRevA.99.013804,PhysRevA.100.063840}, unconventional PB mechanism based on destructive quantum interference between excitation paths has also been studied  \cite{Guan_2017,shi2018tunable}. This unconventional PB could work well with weak mechanical nonlinearity, while the limitation in producing a large mean phonon number still exists. 

In this work, we present an experimentally feasible method for realizing strong PB with a large mean phonon number in a hybrid spin-mechanical system which does not require working in the strong-coupling regime, addressing the two main problems mentioned above. Specifically, our proposal is based on a well-designed hybrid device where a single nitrogen-vacancy (NV) center embedded in a magnetic field gradient couples to a diamond cantilever resonator through the position-dependent Zeeman shift \cite{PhysRevB.79.041302,rabl2010quantum}. With a particular geometric arrangement of nanomagnets \cite{PhysRevLett.121.123604}, the embedded NV center at the equilibrium position senses null first-order magnetic gradient, thus leading to a quadratic coupling between the NV spin and the resonator. In particular, we introduce a periodic drive on the mechanical cantilever to implement the process of mechanical parametric amplification (MPA), which is inspired by a recent work for enhancing spin-phonon coupling \cite{PhysRevLett.125.153602}. By working in the squeezed frame, we show that our implementation enables an { exponential} coupling enhancement with respect to the bare spin-phonon coupling, which differs from the existing schemes based on MPA or optical parametric amplification \cite{PhysRevLett.125.153602,PhysRevLett.114.093602,PhysRevLett.120.093601,PhysRevLett.120.093602,PhysRevA.99.023833,https://doi.org/10.1002/andp.201900220,PhysRevA.100.012339,PhysRevA.100.062501,PhysRevA.101.053826,PhysRevA.102.032601,Wang:20,PhysRevLett.126.023602}. This offers a versatile platform for practical applications in solid-state systems where strong spin-mechanical coupling at the level of two phonons is within reach. 

As the main concern of this work, we propose to improve the performance of PB occurring in our hybrid system with the two-phonon nonlinearity. We show that the { introduction} of MPA allows for a stronger PB compared to the case without MPA, which makes, in principle, the quantum effect of PB as strong as possible. In particular, we demonstrate that the PB is tunable by simply modulating MPA, thereby enabling parameter regions where ultra-strong PB and a large probability for detection of single phonon could coexist even when the system is originally in the weak-coupling regime. Furthermore, despite the lack of direct reference for experimental detection of PB, we analyze a promising method to implementing PB detection through measuring the spin qubit.
 
 { Our investigation on PB provides an alternative route to single-phonon generation, where enhanced mechanical nonlinearity enables a large second-order correlation and thus could facilitate the realization of a high-efficiency single-phonon source and novel single-phonon quantum devices, with extended applications in quantum phononics and phononic quantum networks \cite{Habraken_2012,PhysRevX.8.041027,PhysRevLett.120.213603}. Given the fact that single phonon can be generated by various means, such as phonon sideband transitions of NV centers \cite{PhysRevB.88.064105,PhysRevX.6.041060,PhysRevLett.120.167401}, strong photon-phonon interactions in optomechanical crystals \cite{PhysRevLett.116.234301,hong2017hanbury}, or strong coupling between a superconducting qubit and the phonon modes of an acoustic wave resonator \cite{chu2017quantum}, engineering an efficient single-phonon source may be of particular significance for quantum computing with phonons \cite{PhysRevLett.110.120503,PhysRevX.11.031027}. Furthermore, in analogy to the single-photon sources that have important applications within quantum information science, high-quality single phonon sources may directly determine the development and performance of phonon-based quantum information technologies.
}

\section{Design of the device}\label{sec2}
\subsection{Model and hybrid-system Hamiltonian}
\begin{figure}
	\includegraphics[width=1\columnwidth]{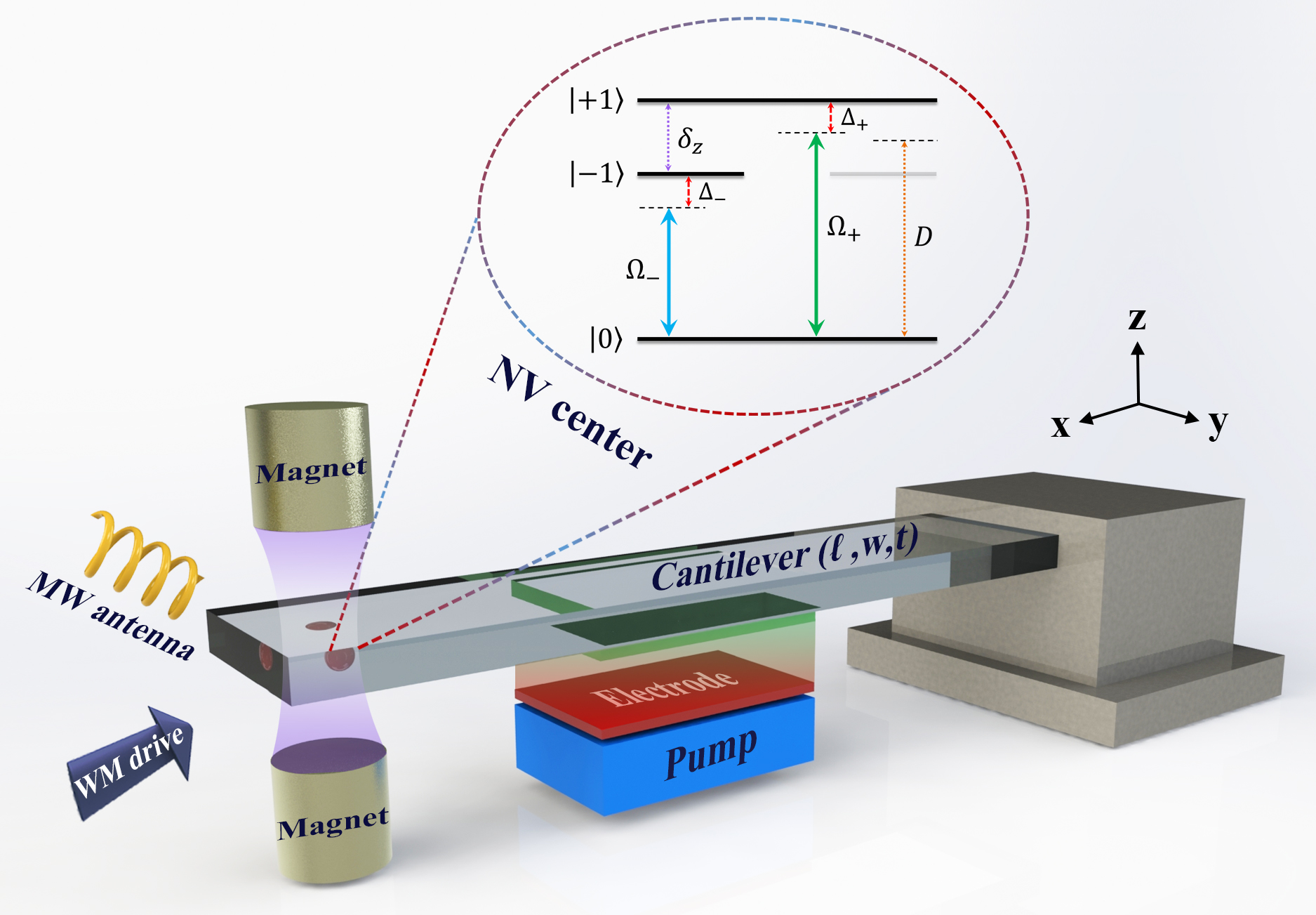}
	\caption{Schematic illustration of the proposed setup for observing strong PB. A diamond cantilever hosting an NV center is placed between two identical nanomagnets, under which a pump source is arranged to electrically modulate its spring constant. A time-varying voltage originated from the tunable oscillating pump is applied, which forms a general capacitor between the two electrodes, one of which is coated on the lower surface of the cantilever and another is placed on top of the pump. A weak mechanical drive is applied to the cantilever. Also, a microwave antenna is used to manipulate the NV electronic spin. Dimensions of each components are not to scale for visual clarity. The inset shows the level diagram of the NV center, whose energy-level transitions between electronic ground states are driven by microwave fields.}
	\label{fig1}
\end{figure}

We consider a hybrid spin-mechanical setup, as schematically illustrated in Fig.~\ref{fig1}, where a single NV center is embedded near the freely vibrating end of a singly-clamped diamond cantilever of dimensions ($\ell,w,t$). Two cylindrical nanomagnets are symmetrically arranged on the two sides of the cantilever to produce strong magnetic field gradient at short distances \cite{PhysRevA.94.053836,cai2017second,PhysRevLett.121.123604,PhysRevA.100.063840}. The electronic spin of the NV center can be coupled to the cantilever resonator based on the fact that the bending motion of the cantilever modulates the local magnetic field sensed by the NV spin \cite{arcizet2011single,PhysRevB.94.214115}.
{ The cantilever is electrically pumped by a periodic drive (under the cantilever) which modulates the mechanical spring constant in time \cite{PhysRevLett.67.699,lemonde2016enhanced,PhysRevLett.125.153602}, generating the process of MPA. In addition, a { weak mechanical drive} is applied to the cantilever for inducing the PB effect. 
	
Following the specific quantization process as in Refs.~\cite{lemonde2016enhanced,PhysRevLett.125.153602}, the Hamiltonian of the cantilever oscillator takes the form
\begin{equation}
\begin{split}
\hat{H}_{\mathrm{mec}}=&\omega_{m} \hat{a}^{\dagger} \hat{a}+\Omega_{p} \cos \left(2 \omega_{p} t\right)\left(\hat{a}^{\dagger}+\hat{a}\right)^{2}\\&+\varepsilon_L\left(\hat{a}^{\dagger}e^{-i\omega_L t}+ \hat{a} e^{i\omega_L t}\right), 
\end{split}
\end{equation} 
where $\hat{a}$ ($\hat{a}^{\dagger}$) is the annihilation (creation) operator for the fundamental vibrational mode of frequency $\omega_{m}$,  $\omega_{p}$ and $\Omega_{p}$ are the pump frequency and amplitude, $\omega_L$ and $\varepsilon_L$ are the frequency and strength of the linear mechanical drive. 
} 


The electronic ground state of the single NV center we studied is a $S=1$ spin triplet with basis states $|m_s\rangle$, where $m_s=0,\pm1$. The ground-state energy-level structure of this NV center is shown in the inset of Fig.~\ref{fig1}. The zero-field splitting between the degenerate sublevels $|m_s=\pm1\rangle$ and $|m_s=0\rangle$ is $D = 2\pi \times 2.87$ GHz. { The strong magnetic field $\textbf{B}$ generated by the nanomagnets at the position of the NV center (see below) causes a Zeeman splitting much larger than the zero-field splitting $D$, which can be canceled by introducing a static magnetic field $\textbf{B}_{\text {static}}$ (in the opposite direction to $\textbf{B}$) \cite{PhysRevB.79.041302}. The magnetic fields of these two parts form a superposed field of magnitude $B_{s}$ such that $\mu_{B} B_{s}\ll D$, thus giving a Zeeman splitting $\delta_z=2 g_{e} \mu_{B} B_{s}$, where $g_{e} \simeq 2$ is the { Land$\acute{\mathrm{e}}$ g-factor of NV centers} and $\mu_{B}=14\ \mathrm{GHz} / \mathrm{T}$ the Bohr magneton.} On this basis, two microwave (MW) fields  polarized in the $x$
direction, $B_{x}^{\pm}(t)=B_{0}^{\pm} \cos \omega_{\pm} t$, are introduced to induce the transitions between states $|\pm1\rangle$ and $|0\rangle$, whose Rabi frequencies $\Omega_{\pm}\equiv g_{e} \mu_{B} B_{0}^{\pm}$ and (red) detunings $\Delta_\pm\equiv D \pm \delta_z/2-\omega_{\pm}$ are supposed to be identical in what follows for simplicity, i.e., $\Omega_{\pm}=\Omega$ and $\Delta_{\pm}=\Delta$.  In a rotating frame with respect to the MW frequencies $\omega_\pm$ and under the rotating-wave approximation (RWA), the Hamiltonian of the single NV center is written as \cite{PhysRevB.79.041302}
\begin{equation}
\hat{H}_{\mathrm{NV}}=\sum_{j=\pm}\Delta|j\rangle\langle j|+\frac{\Omega}{2}(|0\rangle\langle j|+| j\rangle\langle 0|).
\end{equation}

\begin{figure}
	\includegraphics[width=1\columnwidth]{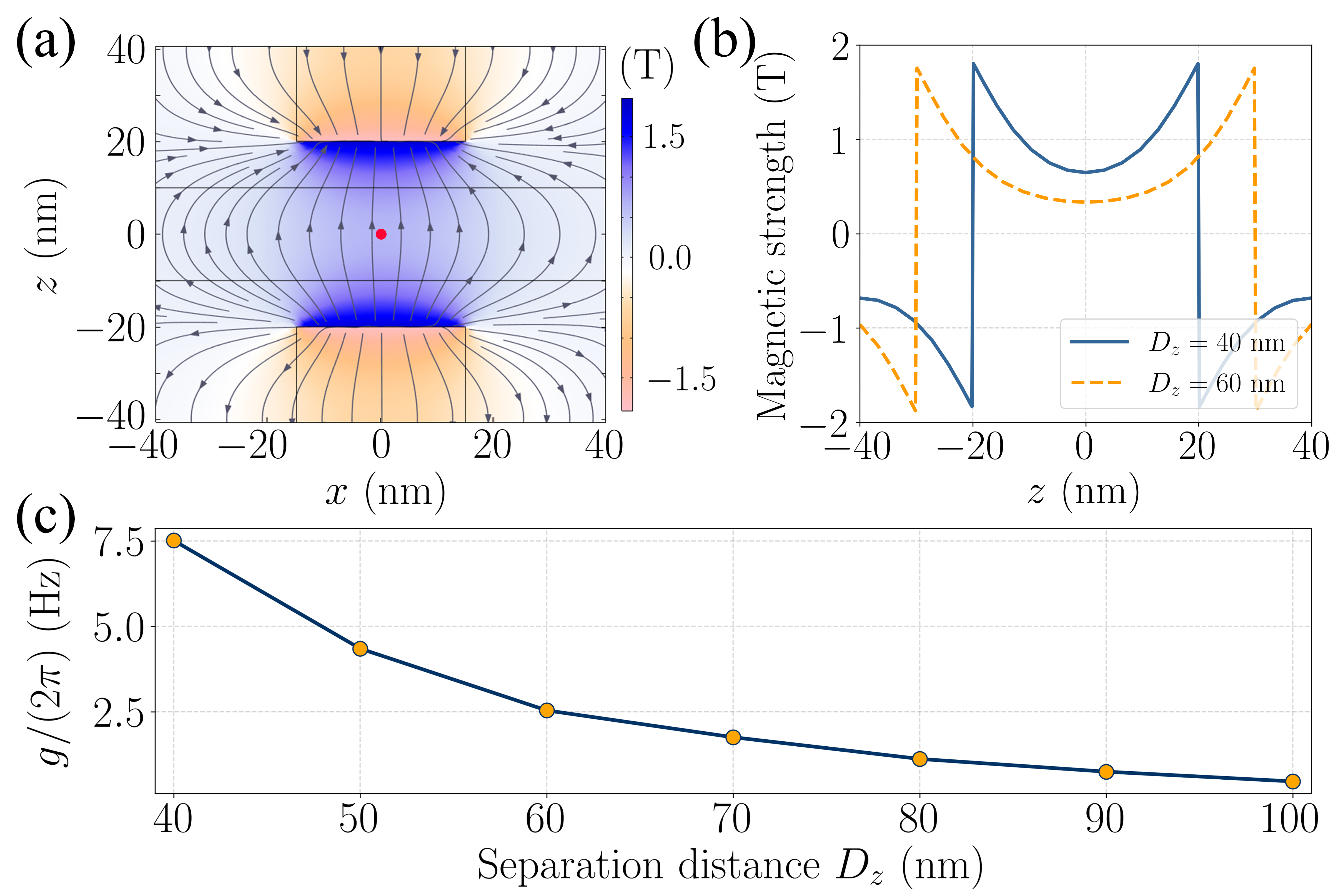}
	\caption{(a) Finite-element simulations of magnetic field lines and strength distribution generated by two cylindrical Dy nanomagnets with a diamond cantilever (hosting an NV center, marked with a red dot at the original point) situated in the middle. Only the $xz$ plane of the spacial magnetic distributions is given here. Two cylindrical magnets have the same diameter of 30 nm and height of 40 nm, which are separated by a gap of $D_z=40$ nm. The dimension of the cantilever is ($\ell,w,t$)=(4,0.1,0.02) $\mu$m. (b) Magnetic strength along the $z$ axis for $x = 0$, for two different magnet separation $D_z$. (c) The calculated two-phonon coupling rate versus the separation distance between the nanomagnets.}
	\label{fig2}
\end{figure}

Next, we turn our attention to the spin-phonon interaction. In principle, the single NV spin surrounded by a magnetic field $\textbf{B}(\textbf{r})$ couples to the mechanical oscillator via the position-dependent Zeeman shift. The corresponding magnetic interaction Hamiltonian is generally written as $\hat{H}_{\mathrm{int}}=\mu_{B} g_{e} \hat{\textbf{S}}\cdot \textbf{B}(\textbf{r}_0)$, where $\hat{\textbf{S}}$ denotes the spin operator and $\textbf{r}_0$ the position of the NV center. Assume that the cantilever resonator (also the NV spin) oscillates only along the $z$ direction, which allows us to expand the Hamiltonian $\hat{H}_{\mathrm{int}}$ up to second order in terms of the cantilever displacement $z$, resulting in $\hat{H}_{\mathrm{int}}\simeq \mu_{B} g_{e} \hat{\textbf{S}} \cdot\left[\partial \mathbf{B} / \partial z(0) \hat{z}+\frac{1}{2} \partial^{2} \mathbf{B} / \partial z^{2}(0) \hat{z}^{2}\right]$. Due to the specific arrangement of nanomagnets, the generated magnetic field yields an extremum at the position of the NV center, thus nullifying the first derivative $\partial \mathbf{B} / \partial z(0)$, that is, the first-order magnetic gradient is equal to zero. This can be verified in Fig.~\ref{fig2}, where we show in Fig.~\ref{fig2}(a) the magnetic field distribution in the $xz$ plane generated by two symmetrically-placed nanomagnets, and the magnetic strength along the $z$ axis for $x = 0$ is given in Fig.~\ref{fig2}(b). Here, the two cylindrical magnets are of the same size, with a diameter of 30 nm and height of 40 nm, and the distance between them is adjustable, ranging from tens to hundreds of nanometers; The magnetic substance is selected as Dy due to a high saturation magnetization up to $\mu_{0} M_{\mathrm{s}}=3.7$ T \cite{PhysRevLett.121.123604}. For the diamond cantilever, its dimension is set as ($\ell,w,t$)=(4,0.1,0.02) $\mu$m in our simulations. { Note that the components of different dimensions as simulated here could be fabricated individually with modern nanofabrication techniques; See Appendix \ref{appenA} for a brief discussion regarding experimental realizations of device fabrication and arrangement.} As seen in Fig.~\ref{fig2}(b), the magnetic strength displays an { extremum} (precisely, a local minimum) at $z=0$ where the NV center is located. Also, it shows that the magnetic strength between the nanomagnets decreases with the increase of the magnet separation $D_z$.

The resulting interaction Hamiltonian $\hat{H}_{\mathrm{int}}$, after eliminating the first-order magnetic gradient, is given by $\hat{H}_{\mathrm{int}}\simeq \frac{1}{2}\mu_{B} g_{e} G  \hat{z} ^{2}  \hat{S}_z $, where $G=\partial^{2} B_z / \partial z^{2}(0)$ represents the second-order magnetic gradient. Note here that the mechanical oscillator only couples to the $z$-component of the NV spin due to null second derivatives of the magnetic field along the $x$ and $y$ directions. Expressing the position operator $\hat{z}$ with $\hat{z}=z_{\mathrm{zpf}}\left(\hat{a}^{\dagger}+\hat{a}\right)$, we end up with
\begin{equation}
\hat{H}_{\mathrm{int}}\simeq  g\left( \hat{a}^{\dagger}+\hat{a}\right)^2  \hat{S}_z,
\end{equation}
where $g=\frac{1}{2}\mu_{B} g_{e} z_{\mathrm{zpf}} ^{2} G $ is the two-phonon coupling rate. Since $g$ is proportional to the square of the zero-point motion of the oscillator $z_{\mathrm{zpf}}$, which ranges from tens to hundreds of femtometers for various systems \cite{ghadimi2018elastic,PhysRevLett.116.147202,weber2016force,singh2014optomechanical,arcizet2011single}, the quadratic coupling strength is intrinsically much smaller than the single-phonon coupling strength ($g_0$) that is proportional to the zero-point fluctuation ($g_0\propto z_{\mathrm{zpf}}$) \cite{PhysRevB.79.041302,rabl2010quantum,PhysRevLett.125.153602,PhysRevA.98.052346,PhysRevApplied.4.044003,PhysRevLett.117.015502}. Also, the magnitude of the second-order magnetic gradient determines linearly the quadratic coupling strength, which is controllable over a large range by adjusting the magnets separation (see below).

To sum up, we obtain the total Hamiltonian for the studied hybrid system 
\begin{equation} \label{H_total}
\begin{split}
\hat{H}_{\mathrm{Total}}=&\hat{H}_{\mathrm{mec}}+\hat{H}_{\mathrm{NV}}+\hat{H}_{\mathrm{int}}\\
=&\omega_{m} \hat{a}^{\dagger} \hat{a}+\sum_{j=\pm}\Delta|j\rangle\langle j|+\frac{\Omega}{2}(|0\rangle\langle j|+| j\rangle\langle 0|)\\
&+g\left( \hat{a}^{\dagger}+\hat{a}\right)^2  \hat{S}_z+\Omega_{p} \cos \left(2 \omega_{p} t\right)\left(\hat{a}^{\dagger}+\hat{a}\right)^{2}
\\
&+\varepsilon_L\left(\hat{a}^{\dagger}e^{-i\omega_L t}+ \hat{a} e^{i\omega_L t}\right).
\end{split}
\end{equation}
To diagonalize the spin-only part (i.e., $\hat{H}_{\mathrm{NV}}$), we then transform $\hat{H}_{\mathrm{Total}}$ to a dressed-state basis 
spanned by $\{|\mathcal{D}\rangle=(|+1\rangle-|-1\rangle)/\sqrt{2},|\mathcal{G}\rangle=\cos \theta|0\rangle-\sin \theta|\mathcal{B}\rangle,|\mathcal{E}\rangle=\cos \theta|\mathcal{B}\rangle+\sin \theta|0\rangle\}$, with $|\mathcal{B}\rangle=(|+1\rangle+|-1\rangle) / \sqrt{2}$  and  $\tan (2 \theta)=\sqrt{2} \Omega / \Delta$. The resulting system Hamiltonian in a rotating frame with respect to $\omega_p$, under the RWA, is simplified as (see Appendix \ref{appenC} for a detailed derivation)
\begin{equation}\label{JC_H}
\begin{split}
\hat{H}_{\mathrm{Total}}\simeq &\delta_{m} \hat{a}^{\dagger} \hat{a}+\delta_{ed}\hat{\sigma}_{+}\hat{\sigma}_{-}+ g\left( \hat{a}^{\dagger2}\hat{\sigma}_{-}+\hat{a}^{2}\hat{\sigma}_{+}\right)\\
&+\frac{\Omega_{p}}{2} \left(\hat{a}^{\dagger 2}+\hat{a}^{2}\right)+\varepsilon_L\left(\hat{a}^{\dagger}e^{-i\delta_L t}+ \hat{a} e^{i\delta_L t}\right).
\end{split}
\end{equation}
where $\delta_{m(L)}=\omega_{m(L)}-\omega_{p}$, $\delta_{ed}=\omega_{ed}-2\omega_{p}$, and $\hat{\sigma}_{+} =\hat{\sigma}_{-}^{\dagger}\equiv|\mathcal{E}\rangle\langle \mathcal{D}|$. Notably, Eq.~(\ref{JC_H}) has the form of a two-phonon Jaynes-Cummings (JC) Hamiltonian in the presence of both { linear and nonlinear (two-phonon)} mechanical drive, which suggests degenerate two-phonon exchange between the mechanical mode and an effective TLS. 
 
\subsection{Estimation of the two-phonon coupling rates}
So far, we have derived the driven, two-phonon JC Hamiltonian for the hybrid spin-mechanical system. In this subsection, we proceed to estimate the experimentally achievable two-phonon coupling rate as well as other system parameters. Based on the numerical simulations in Figs.~\ref{fig2}(a) and \ref{fig2}(b), it is straightforward to acquire the values of the second-order magnetic gradient $G=\partial^{2} B_z / \partial z^{2}(0)$ for different magnet separations. Then, the two-phonon coupling $g$ depends directly on the value of the zero-point fluctuation of the oscillator $z_{\mathrm{zpf}}=\sqrt{\hbar / 2 M \omega_{m}}$, according to $g=\frac{1}{2}\mu_{B} g_{e} z_{\mathrm{zpf}} ^{2} G $. The fundamental frequency $\omega_{m}$ and effective mass $M$ of the oscillator are estimated as $\omega_{m} \approx 3.516 \times\left(t / \ell ^{2}\right) \sqrt{E / 12 \rho}\approx 2\pi \times 3.8$ MHz and $ M = \rho \ell w t /4 \approx 7 \times 10^{-18} $ kg, with the Young’s modulus and mass density of diamond being $E \approx 1.22 \times 10^{12} \mathrm{~Pa}$ and  $\rho \approx 3.52 \times 10^{3} \mathrm{~kg/m^3}$, respectively, which gives $z_{\mathrm{zpf}}\approx563$ fm. In Fig.~\ref{fig2}(c) we plot the calculated two-phonon coupling rate $g$ versus the magnet separation $D_z$ ranging from 40 to 100 nm. It shows that a small magnet separation is desirable for achieving a large coupling rate. Despite this, we will take hereafter a relatively large value of $D_z$ from the perspective that a larger magnet separation would be more beneficial to actual operation in experiments. As an example, we take $D_z=80$ nm, giving $G\approx7.9\times 10^{14} \mathrm{~T/m^2}$  and therefore $g\approx 2\pi \times 1.1$ Hz. { A small misalignment in the field orientation due to imperfectly positioned magnets could affect slightly the second-order magnetic gradient and therefore the two-phonon coupling rate, leading to a relative error of less than $2\%$ in the presence of a $10^\circ$ misalignment (see Appendix \ref{appenB} for details).}

\subsection{MPA in the squeezed frame}
\begin{figure}
	\includegraphics[width=1\columnwidth]{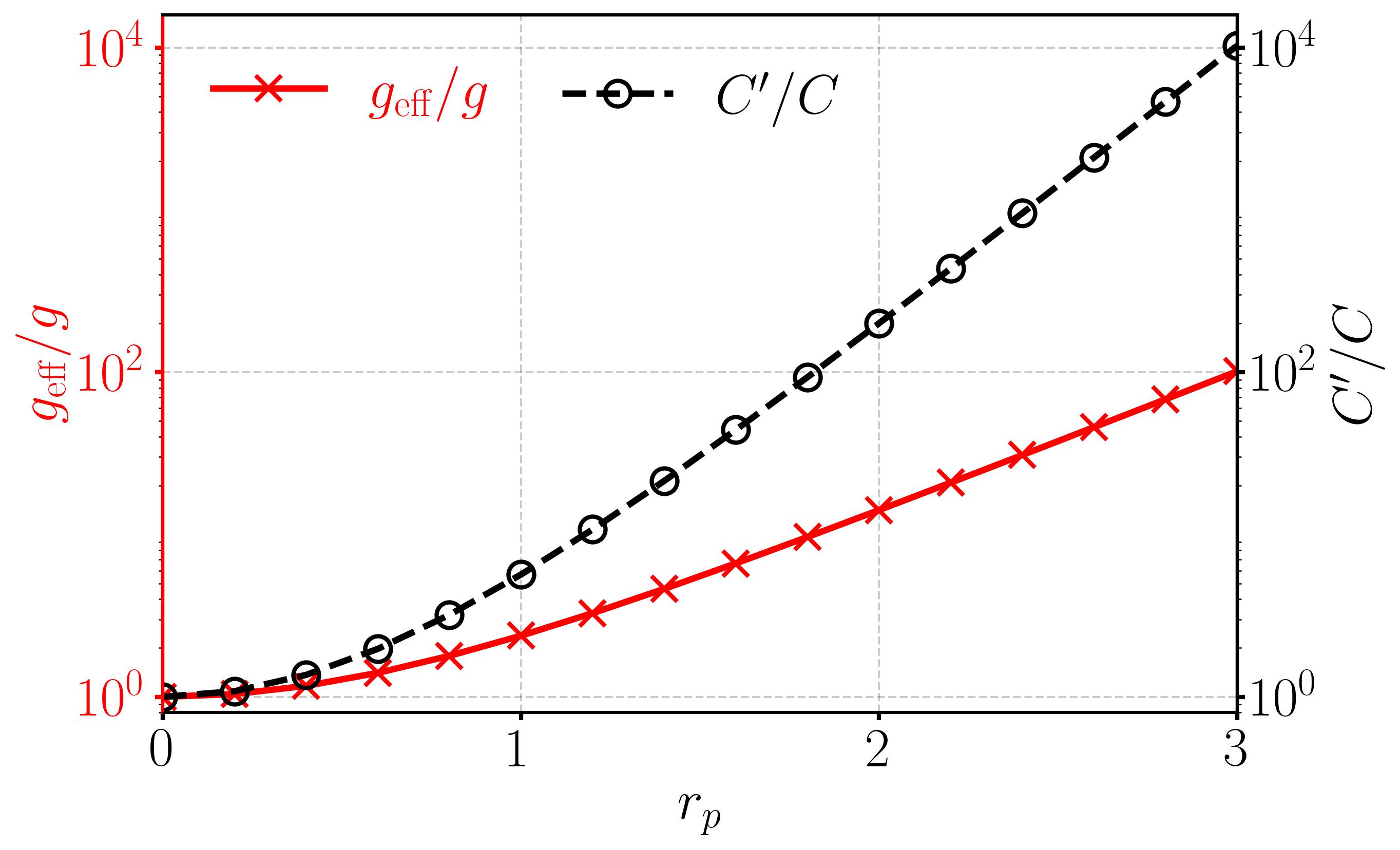}
	\caption{Spin-mechanical coupling enhancement $g_{\mathrm{eff}}/g$ and cooperativity enhancement $C'/C$ versus the squeezing parameter $r_p$.}
	\label{fig3}
\end{figure}

The time-dependent pump on the mechanical oscillator introduces nonlinear drive to the mechanical mode, producing the effect of degenerate parametric amplification. Specifically, by performing a unitary squeezing transformation $\hat{U}_{s}=\exp \left[r_p\left(\hat{a}^{2}-\hat{a}^{\dagger 2}\right) / 2\right]$ \cite{PhysRevLett.120.093602,PhysRevA.101.053826,PhysRevA.102.032601,PhysRevLett.126.023602}, where the squeezing parameter $r_p$ is defined via $r_{p}=(1 / 2) \operatorname{arctanh}\left(\Omega_{p} / \delta_{m}\right)$, the system Hamiltonian (\ref{JC_H}) can be transformed to the squeezed frame, yielding (see Appendix \ref{appenD})
\begin{equation}\label{H_squ}
\begin{split}
\hat{H}_{\mathrm{Total}}^{\mathrm{S}}\simeq &\delta_{s} \hat{a}_{s}^{\dagger} \hat{a}_{s}+\delta_{ed}\hat{\sigma}_{+}\hat{\sigma}_{-} +g_{\mathrm{eff}}(\hat{a}_{s}^{\dagger2}\hat{\sigma}_{-}+\hat{a}_{s}^{2}\hat{\sigma}_{+})\\
&+\varepsilon_L'\left(\hat{a}_{s}^{\dagger}e^{-i\delta_L t}+\hat{a}_{s}e^{i\delta_L t}\right),
\end{split}
\end{equation}
where $\delta_{s}=\delta_{m}/\cosh(2r_p)$ is the squeezed-oscillator frequency, $g_{\mathrm{eff}}=g\cosh^2(r_p)$ and $\varepsilon_L'=\varepsilon_L\cosh(r_p)$ are the effective two-phonon spin-mechanical coupling strength and { linear} mechanical driving strength, respectively. The { exponential coupling enhancement with respect to the original coupling strength, $g_{\mathrm{eff}}/g=\cosh^2(r_p)$}, is visualized in Fig.~\ref{fig3}, where one sees that $g_{\mathrm{eff}}$ can be two orders of magnitude larger than $g$ for a modest squeezing parameter $r_p=3$. It should be mentioned that the scheme proposed here is distinct from the exponential enhancement associated with the single-phonon (photon) spin-resonator interaction using mechanical (optical) parametric amplification \cite{PhysRevLett.125.153602,PhysRevLett.114.093602,PhysRevLett.120.093601,PhysRevLett.120.093602,PhysRevA.99.023833,https://doi.org/10.1002/andp.201900220,PhysRevA.100.012339,PhysRevA.100.062501,PhysRevA.101.053826,PhysRevA.102.032601,Wang:20,PhysRevLett.126.023602}. 

Utilizing MPA to enhance the spin-mechanical coupling also amplifies the mechanical noise inevitably, which could break down any nonclassical behavior in the system due to the amplified dissipation. { An effective strategy against this adverse effect is to use the squeezed-vacuum-reservoir technique, which has been studied extensively in recent reports \cite{PhysRevLett.125.153602,PhysRevLett.114.093602,PhysRevLett.120.093601,PhysRevLett.120.093602,PhysRevA.99.023833,https://doi.org/10.1002/andp.201900220,PhysRevA.100.012339,PhysRevA.100.062501,PhysRevA.101.053826,PhysRevA.102.032601,PhysRevLett.126.023602}. To be specific, by integrating an auxiliary, broadband microwave resonator to the original hybrid system, which acts as an engineered squeezed-vacuum reservoir, the squeezing-enhanced mechanical noise could be effectively suppressed, ensuring that the squeezed mechanical mode equivalently interacts with a thermal vacuum reservoir (see Appendix \ref{appenF} for details).} The dynamics of the system is thus governed by the standard Lindblad master equation which in the squeezed frame takes the form \cite{PhysRevLett.125.153602}
\begin{equation} \label{mas equ}
\dot{\hat{\rho}}=i\left[\hat{\rho}, \hat{H}_{\mathrm{Total}}^{\mathrm{S}}\right]+\gamma_{m_{\mathrm{eff}}} \mathcal{D}[\hat{a}_{s}] \hat{\rho}+\gamma_{z} \mathcal{D}\left[\hat{\sigma}_{+}\hat{\sigma}_{-}\right] \hat{\rho},
\end{equation}
where $\mathcal{D}[\hat{o}] \hat{\rho}=\hat{o} \hat{\rho} \hat{o}^{\dagger}-\hat{o}^{\dagger} \hat{o} \hat{\rho} / 2-\hat{\rho} \hat{o}^{\dagger} \hat{o} / 2$, $\gamma_{m_{\mathrm{eff}}}$ is the engineered effective mechanical dissipation rate as a consequence of the coupling between the mechanical mode and the auxiliary reservoir, and $\gamma_{z}$ is the dephasing rate of the spin qubit. The quality factor of the oscillator $Q$ is associated with $\gamma_{m_{\mathrm{eff}}}$ via the relation $Q=n_{\mathrm{th}}\omega_m/\gamma_{m_{\mathrm{eff}}}$, with $n_{\mathrm{th}}=\left[\exp \left(\hbar \omega_{m} / k_{B} T\right)-1\right]^{-1}$ being the
average number of thermal phonons in the oscillator at the temperature $T$.  In the present work the environmental temperature is assumed to be about tens of millikelvin \cite{tsaturyan2017ultracoherent}, resulting in tens of mean thermal phonon number for the mechanical frequency of tens of MHz. When it comes to the NV center, its dephasing time, given by $T_{2}=1/\gamma_z$, is assumed to be on the order of milliseconds in our calculations. While the spin properties of shallow NV centers are not as good as those for NV centers in bulk diamond \cite{bar2013solid,de2010universal,kolkowitz2012coherent}, many experimental efforts have been devoted to improve the coherence properties of shallow NV centers \cite{mamin2013nanoscale,ishikawa2012optical,ohashi2013negatively,doi:10.1063/1.4748280,PhysRevLett.113.027602,PhysRevLett.113.027602}, which make the extension of coherence time possible with the development and exploration of future experimental techniques. For clarity, we also evaluate the effect of spin dephasing on the PB performance by simulating both steady-state second-order correlation function and mean phonon number with varying spin dephasing rate over a large range (see below).


The master equation (\ref{mas equ}) gives an effective cooperativity $C'=g_{\mathrm{eff}}^2/(\gamma_{m_{\mathrm{eff}}} \gamma_{z})$, which can be strengthened significantly with increasing the squeezing parameter $r_p$, as shown in Fig.~\ref{fig3}. Here, the cooperativity enhancement is approximated as $C'/C\approx\cosh^4(r_p)$. This result is justified for the case where the effective mechanical dissipation rate is almost unchanged after introducing MPA. The enhancement in the effective cooperativity provides great potentials for various applications in quantum information technologies. In the following, we explore the possibilities of using the designed device to induce strong PB and investigate the statistical characteristics of phonons. As we will show, both analytically and numerically, the enhanced cooperativity contributes essentially to the observation of strong PB in the hybrid system that works in the weak-coupling regime.

\section{Analytical description of the enhanced PB}\label{sec3}
\begin{figure}
	\includegraphics[width=1\columnwidth]{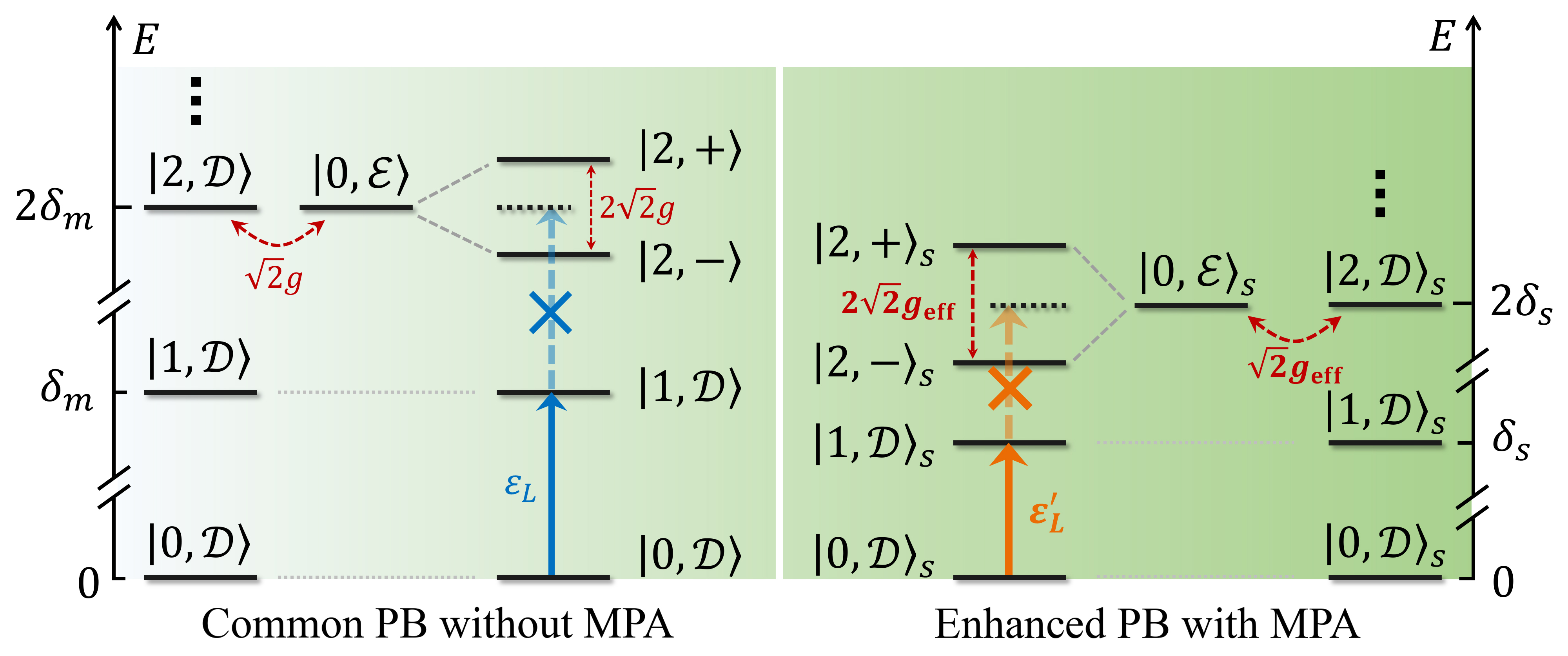}
	\caption{Comparison of the energy-level diagrams explaining the origin of the enhanced PB in our weakly coupled hybrid system: mechanical parametric drive amplifies the anharmonic spacing, thereby strongly suppressing the excitation of the second phonon.}
	\label{fig4}
\end{figure}

In the absence of the mechanical parametric drive, PB can appear in the hybrid system with a two-phonon nonlinearity of JC type. Specifically, for the system initially prepared in its ground state, i.e.,  $|0,\mathcal{D}\rangle$, the first phonon of the oscillator can be easily generated under a resonant mechanical drive $\varepsilon_L$, via the transition $|0,\mathcal{D}\rangle\rightarrow|1,\mathcal{D}\rangle$, where $|1,\mathcal{D}\rangle$ represents the first excited state of the system. The two-phonon spin-mechanical interaction dresses the two bare states $|2,\mathcal{D}\rangle$ and $|0,\mathcal{E}\rangle$, resulting in second excited states of the system being superposition forms $|2,\pm\rangle\equiv\left( |2,\mathcal{D}\rangle\pm|0,\mathcal{E}\rangle\right) /\sqrt{2}$, with an energy splitting $2\sqrt{2}g$ (see the left panel in Fig.~\ref{fig4}). For a sufficiently strong coupling rate $g$, the transition $|1,\mathcal{D}\rangle\rightarrow|2,\pm\rangle$ is detuned and, thus, suppressed for $g\gg\gamma_{m_\mathrm{{eff}}}$. This gives a clear signature of single PB:
once a phonon is coupled to the system, it suppresses the probability of coupling a second phonon with the same frequency. Intrinsically, the PB predicted here is a consequence of the energy-structure anharmonicity, which requires a strong nonlinear coupling $g$ to ensure the system being in the strong-coupling regime.

Perhaps more interesting is the ability of our scheme to access strong PB even in a weakly coupled system by taking advantage of MPA. That is, the scheme here is free from the restriction of strong coupling which is generally prerequisite in reported schemes \cite{PhysRevA.100.063840,PhysRevA.99.013804,PhysRevA.93.063861,PhysRevA.96.013861,PhysRevA.94.063853,PhysRevA.95.053844,PhysRevA.98.023819,PhysRevLett.110.193602,PhysRevA.82.032101}. This characteristic originates from the fact that the mechanical parametric drive amplifies the anharmonicity of the eigenenergy spectrum, as depicted in the right panel of Fig.~\ref{fig4}, thus suppressing more drastically the excitation of the second phonon. To confirm this intuitive picture, in the following, we describe quantitatively the PB by studying the phonon-number distribution and the phonon correlation functions.

\subsection{Criteria of PB}
We start by introducing two criteria for PB. The first criterion is based on a comparison of the phonon-number distribution with standard Poissonian distribution. Specifically, when $n$-PB occurs, the phonon-number distribution $P(m)$ (with normalization $\sum_{m=0}^{\infty} P(m)=1$) satisfies \cite{PhysRevLett.118.133604,PhysRevLett.121.153601}
\begin{equation} \label{criterion 1}
\begin{split}
(i) \quad & P(m)<\mathcal{P}(m) \quad \text {for} \ m>n,\\
(ii) \quad & P(n) \ge \mathcal{P}(n),
\end{split}
\end{equation}
where $\mathcal{P}(m)=\langle\hat{m}\rangle^{m} e^{ -\langle\hat{m}\rangle} /m ! $ is the Poissonian distribution with the same average phonon number $\langle\hat{m}\rangle$ as the mechanical oscillator. This indicates a sub-Poissonian phonon-number statistics for $(n + 1)$ phonons with the simultaneous super-Poissonian statistics of the first $n$ phonons. Also, the relative deviation of a given phonon-number distribution from the corresponding Poissonian distribution is given by $[P(m)-\mathcal{P}(m)] / \mathcal{P}(m)$.

The second criterion is based on calculating the phonon correlation functions that characterize the statistical properties of phonons. The normalized  equal-time $\mu$th-order correlation function is defined as
\begin{equation} 
g^{(\mu)}(t, 0)=\frac{\left\langle \hat{a}^{\dagger\mu}(t) \hat{a}^{\mu}(t)\right\rangle}{\left\langle \hat{a}^{\dagger}(t) \hat{a}(t)\right\rangle^{\mu}},
\end{equation}  
which is related to the probability of simultaneously measuring $\mu$ phonons. In the steady state, the equal-time $\mu$th-order correlation function becomes  $g^{(\mu)}(0)_{ss} = \text{lim}_{t\rightarrow\infty}g^{(\mu)}(t, 0)$. Note that the larger value of $g^{(\mu)}(0)_{ss} > 1$, the higher probability of $\mu$-phonon bunching; The smaller value of $g^{(\mu)}(0)_{ss} < 1$, the higher probability of $\mu$-phonon antibunching. Particularly, the steady-state, equal-time second-order correlation function is given by
\begin{equation} \label{g2}
g^{(2)}(0)_{ss}=\frac{\left\langle \hat{a}^{\dagger 2} \hat{a}^{2}\right\rangle}{\left\langle \hat{a}^{\dagger} \hat{a}\right\rangle^{2}}.
\end{equation}  
In the case of a finite-time delay between two phonon detection, one can apply the steady-state, delayed-time second-order correlation function
\begin{equation}\label{delayed g2}
g^{(2)}(\tau)_{ss}=\text{lim}_{t\rightarrow\infty}\frac{\left\langle \hat{a}^{\dagger}(t) \hat{a}^{\dagger}(t+\tau) \hat{a}(t+\tau) \hat{a}(t)\right\rangle}{\left\langle \hat{a}^{\dagger}(t) \hat{a}(t)\right\rangle\left\langle \hat{a}^{\dagger}(t+\tau) \hat{a}(t+\tau)\right\rangle}.
\end{equation} 

The criterion for $n$-phonon blockade with respect to the phonon-number distribution, given in Eq.~(\ref{criterion 1}), can be translated into the following conditions \cite{PhysRevLett.118.133604,PhysRevLett.121.153601}
\begin{equation} \label{criterion 2}
\begin{split}
(i)  \quad &g^{(n+1)}(0)_{ss}<\exp (-\langle\hat{m}\rangle),\\
(ii) \quad &g^{(n)}(0)_{ss} \geq \exp (-\langle\hat{m}\rangle)+\langle\hat{m}\rangle \cdot g^{(n+1)}(0)_{ss}.
\end{split}
\end{equation}
For single PB with $n=1$, we obtain

\begin{align}
\label{criterion 2a for 1PB} (i)  \quad &g^{(2)}(0)_{ss}<\exp (-\langle\hat{m}\rangle)\equiv f,\\
\label{criterion 2b for 1PB} (ii) \quad &g^{(1)}(0)_{ss} \geq \exp (-\langle\hat{m}\rangle)+\langle\hat{m}\rangle \cdot g^{(n+1)}(0)_{ss}\equiv f^{(1)}.
\end{align}
For the system with a { weak linear driving of the mechanical resonator}, the mean phonon number is very small, i.e., $\langle\hat{m}\rangle\ll1$, which approximatively gives $f\rightarrow1$ and $f^{(1)}\rightarrow1$. Then the condition simplifies to the usual
criterion of single PB, $g^{(2)}(0)_{ss}<1$. The two criteria, given respectively in Eqs.~(\ref{criterion 1}) and (\ref{criterion 2a for 1PB})-(\ref{criterion 2b for 1PB}), will facilitate us to identify the occurrence of single PB in the studied system.

\subsection{Analytical solution of the second-order correlation function} \label{Ana B}
To analyze the steady state of our system and obtain the analytical expression of the second-order correlation function, we introduce an effective non-Hermitian Hamiltonian involving the system dissipation
\begin{equation}\label{H_nH}
\hat{H}_{\mathrm{nH}}^{\mathrm{S}}= \hat{H}_{\mathrm{eff}}^{\mathrm{S}}-i \frac{\gamma_{m_\mathrm{{eff}}}}{2}\hat{a}_{s}^{\dagger}\hat{a}_{s}-i \frac{\gamma_{\mathrm{z}}}{2}|\mathcal{E}\rangle\langle \mathcal{E}|,
\end{equation}
where
\begin{equation}
\hat{H}_{\mathrm{eff}}^{\mathrm{S}}=\delta\left(\hat{a}_{s}^{\dagger}\hat{a}_{s}\!+\!2 |\mathcal{E}\rangle\langle \mathcal{E}|  \right) \!+\!g_{\mathrm{eff}}(\hat{a}_{s}^{\dagger2}\hat{\sigma}_{-}\!+\!\hat{a}_{s}^{2}\hat{\sigma}_{+})\!+\!\varepsilon_L'\left(\hat{a}_{s}^{\dagger}\!+\!\hat{a}_{s}\right)
\end{equation}
is a reformulation of Eq.~(\ref{H_squ}) after assuming $\delta_{ed}=2\delta_s$ and $\delta_s-\delta_L=\delta$ with $\delta$ being a small laser-drive detuning. In the weak-driving regime $(\varepsilon_L'\ll g_{\mathrm{eff}})$, by truncating the infinite-dimensional Hilbert space to the two-phonon excitation subspace, the wave
function of the system could be expressed with the ansatz
\begin{equation} \label{psi}
|\psi\rangle_\mathrm{s}=C_{0d}|0,\mathcal{D}\rangle_\mathrm{s}+C_{1d}|1,\mathcal{D}\rangle_\mathrm{s}+C_{2d}|2,\mathcal{D}\rangle_\mathrm{s}+C_{0e}|0,\mathcal{E}\rangle_\mathrm{s}, 
\end{equation}
with the coefficients being probability amplitudes of the basis states. Substituting the state $|\psi\rangle_\mathrm{s}$ and the Hamiltonian $\hat{H}_{\mathrm{nH}}^{\mathrm{S}}$ to the Schr$\ddot{\mathrm{o}}$dinger equation $i|\dot{\psi}\rangle_\mathrm{s}=\hat{H}_{\mathrm{nH}}^{\mathrm{S}}|\psi\rangle_\mathrm{s}$, we obtain the following equations of motion for the probability amplitudes
\begin{equation}
\begin{split}
&\dot{C}_{1 d}=-i \varepsilon_{L}' C_{0d}-i\left(\delta-i\frac{\gamma_{m_\mathrm{{eff}}}}{2}\right) C_{1 d}-i\sqrt{2} \varepsilon_{L}' C_{2 d}, \\
&\dot{C}_{2 d}=-i\sqrt{2} \varepsilon_{L}' C_{1d}-i\left(2\delta-i\gamma_{m_\mathrm{{eff}}}\right) C_{2d}-i\sqrt{2} g_{\mathrm{eff}} C_{0 e}, \\
&\dot{C}_{0 e}=-i\sqrt{2} g_{\mathrm{eff}} C_{2d}-i\left(2 \delta-i\frac{\gamma_{\mathrm{z}}}{2}\right)  C_{0 e}.
\end{split}
\end{equation}

By solving the above equations, the steady-state solutions of the subsystem are readily achieved (as given detailedly in Appendix \ref{appenG}). Then, we obtain the analytical expression of the equal-time second-order correlation function
\begin{equation} \label{g2(0)ss}
g^{(2)}(0)_{ss}\simeq\frac{\left[4\delta^2+(\gamma_{\mathrm{z}}/2)^2\right]\left[\delta^2+\varepsilon_{L}^{\prime 2} +(\gamma_{m_\mathrm{{eff}}}/2)^2\right]}{4\delta^4+\delta^2\zeta+\left[g_{\mathrm{eff}}^2+(\gamma_{\mathrm{z}}\gamma_{m_\mathrm{{eff}}}/4) \right]^2  }
\end{equation}
with $\zeta=(\gamma_{\mathrm{z}}/2)^2+\gamma_{m_\mathrm{{eff}}}^2-4g_{\mathrm{eff}}^2$. As seen in Eq.~(\ref{g2(0)ss}), the value of $g^{(2)}(0)_{ss}$ depends directly on $\delta$: for $\delta=0$, $g^{(2)}(0)_{ss}$ reaches its minimum
\begin{equation} \label{g2(0)ss_min}
g^{(2)}(0)_{ss,\mathrm{min}}=\frac{\gamma_{\mathrm{z}}^2\left(\gamma_{m_\mathrm{{eff}}}^2 +4\varepsilon_{L}^{\prime 2}\right)}{\left(\gamma_{m_\mathrm{{eff}}}\gamma_{\mathrm{z}}+ 4g_{\mathrm{eff}}^2\right)^2  }.
\end{equation}
This implies that strong single PB (or, two-phonon antibunching effect) could occur when the mechanical mode is resonantly driven. Further, for $\varepsilon_{L}'\ll \gamma_{m_\mathrm{{eff}}}$, by expressing $g_{\mathrm{eff}}^2/(\gamma_{m_\mathrm{{eff}}}\gamma_{\mathrm{z}})$ as $C'$ in Eq.~(\ref{g2(0)ss_min}), we have
\begin{equation}
g^{(2)}(0)_{ss,\mathrm{min}}\sim\frac{1}{\left(1+ 4C'\right)^2}.
\end{equation}
This explicitly shows that the remarkably enhanced cooperativity in our scheme could make the quantum effect of PB as strong as possible.

\begin{figure*}
	\includegraphics[width=2\columnwidth]{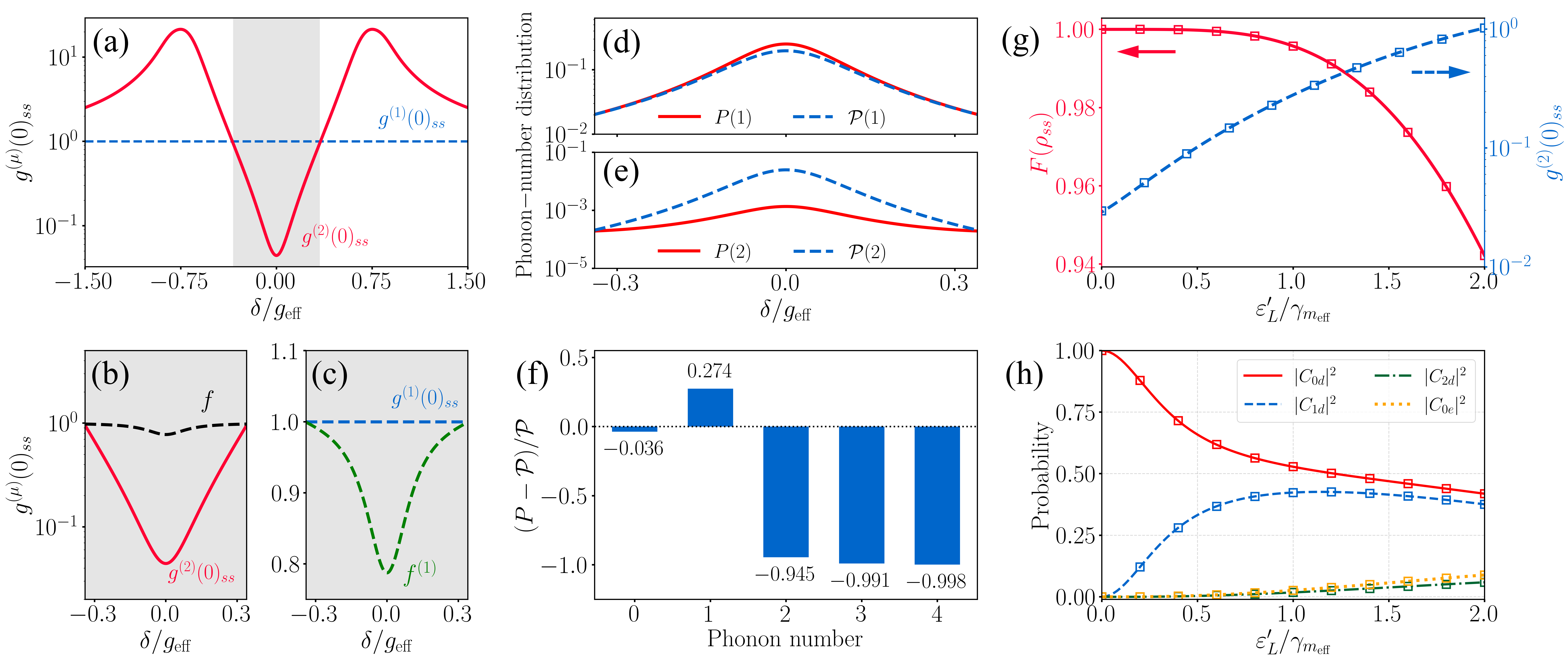}
	\caption{(a) Steady-state equal-time correlation function $g^{(\mu)}(0)_{ss}$ versus driving detuning $\delta/g_{\mathrm{eff}}$. Single PB occurs in the region marked with gray background due to $g^{(2)}(0)_{ss}<f$, as shown in  (b), and $g^{(1)}(0)_{ss}>f^{(1)}$, as depicted in (c), which fulfill the criteria given in Eqs.~(\ref{criterion 2a for 1PB}) and (\ref{criterion 2b for 1PB}), respectively. Also, single PB can be recognized from (d)-(e) the phonon-number distribution and (f) its deviation to the standard Poisson distribution. For $\delta=0$, the single-phonon probability is enhanced as $P(1) > \mathcal{P}(1)$, while the two-phonon probability is suppressed as $P(2) < \mathcal{P}(2)$, which satisfy the criterion given in Eq.(\ref{criterion 1}) for $n=1$. (g) State-truncation fidelity $F\left(\rho_{ss}\right)$ defined in Eq.~(\ref{fidelity}) and second-order correlation function $g^{(2)}(0)_{ss}$ versus the mechanical driving strength $\varepsilon_L'/\gamma_{m_\mathrm{{eff}}}$ for $\delta=0$. (h) Probabilities of the four basis states in the two-phonon excitation subspace, given in Eq.~(\ref{psi}), versus $\varepsilon_L'/\gamma_{m_\mathrm{{eff}}}$. Note that the hollow-square marks in (g) and (h) are obtained by solving numerically the master equation with respect to the full Hamiltonian given in Eq.~(\ref{B1}). The hybrid system starts with its ground state $|0,\mathcal{D}\rangle_\mathrm{s}$. Parameters used here are: (a-h) $r_p=2$, $g=2\pi \times 1.1$  Hz, $\gamma_z=2\pi\times10$ Hz, $n_{\mathrm{th}}=54$, and $\gamma_{m_\mathrm{{eff}}}=2g$; (a-f) $\varepsilon_L'=g_\mathrm{{eff}}/20$; (g-h) $\delta=0$ and $\delta_{ed}=1\times10^{3}g_\mathrm{{eff}}$. }
	\label{fig5}
\end{figure*}

\section{Numerical description of the enhanced PB}
\subsection{PB in the weak-coupling regime}
To confirm the analytical results, we now numerically study the full quantum dynamics of the system. The numerical simulations were performed by solving the master equation (\ref{mas equ}) using the \textit{Quantum Toolbox
in Python} (QuTiP) \cite{JOHANSSON20131234,JOHANSSON20121760}. We consider the system with weak bare coupling: $\gamma_{m_\mathrm{{eff}}}=2g$ \cite{PhysRevLett.117.015502}. Fig.~\ref{fig5}(a) depicts the first- and second-order correlation functions versus the driving detuning $\delta/g_{\mathrm{eff}}$. For $\delta/g_{\mathrm{eff}}$ fluctuating within a small range, the value of $g^{(2)}(0)_{ss}$ keeps below 1, corresponding to the region where PB occurs. In particular, it is shown that $g^{(2)}(0)_{ss}$ has a dip at $\delta=0$ (i.e., a resonant drive), which is consistent with the analytical description above. We also see that, at $\delta=0$, $g^{(2)}(0)_{ss}$ is smaller than $f$ defined in the criterion given in Eq.~(\ref{criterion 2a for 1PB}), while  $g^{(1)}(0)_{ss}$ is greater than $f^{(1)}$ defined in the criterion given in Eq.~(\ref{criterion 2b for 1PB}), as shown in Figs.~\ref{fig5}(b) and \ref{fig5}(c), respectively. Moreover, by comparing the phonon-number distribution $P(m)$, calculated numerically via $P(m)=\left\langle m\left|\hat{\rho}_{ss}\right| m\right\rangle$ with $\hat{\rho}_{ss}$ the steady-state solutions of the master equation, with the Poisson distribution $\mathcal{P}(m)$, we find that the single-phonon probability is enhanced as $P(1) > \mathcal{P}(1)$, while the two-phonon probability is suppressed as $P(2) < \mathcal{P}(2)$, as depicted in Figs.~\ref{fig5}(d) and \ref{fig5}(e), respectively. Fig \ref{fig5}(f) shows the deviations of the phonon distribution to the standard Poisson distribution with the same mean phonon number when $\delta=0$, which characterizes the second-order sub-Poissonian statistics (i.e., two-phonon antibunching). The above results provide sufficient and unambiguous evidence for single PB.

For a weak mechanical drive $\varepsilon_L'$, the
system is well truncated into a two-phonon excitation subspace spanned by four basis states given in Eq.~(\ref{psi}). However, this approximation becomes invalid when higher levels are excited in the case of a relatively strong drive. Thus, in order to estimate the effect of state truncation on the quality of PB, we introduce a fidelity describing the sum of the steady-state probabilities of the four basis states \cite{PhysRevA.87.023809,PhysRevA.100.063840,PhysRevA.93.063861}, given by
	\begin{equation} \label{fidelity}
F\left(\rho_{ss}\right)=\left|C_{0d}\right|^{2}+\left|C_{1d}\right|^{2}+\left|C_{2d}\right|^{2}+\left|C_{0e}\right|^{2}.
\end{equation}
In Fig.~\ref{fig5}(g) we plot the fidelity $F\left(\rho_{s s}\right)$ versus the driving strength $\varepsilon_L'/\gamma_{m_\mathrm{{eff}}}$. For $\varepsilon_L'/\gamma_{m_\mathrm{{eff}}}<0.5$, the value of $F\left(\rho_{s s}\right)$ keeps close to 1, whereas further increasing the driving strength leads to a rapid decrease of $F\left(\rho_{s s}\right)$. In this case, the two-phonon (or even multi-phonon) state will be populated slightly, as seen in Fig.~\ref{fig5}(h), and correspondingly, the effect of PB will be spoiled, resulting in an increased $g^{(2)}(0)_{ss}$ (see the dashed curve in Fig.~\ref{fig5}(g)). Thus, we conclude that the occurrence of strong PB requires the mechanical driving strength to be as small as possible to suppress the two-phonon excitation. Nonetheless, a large single-phonon probability determining the efficiency of single-phonon emission is also crucial for practical applications such as the single-phonon source.

\subsection{Enhancing PB by modulating MPA}
\begin{figure}
	\includegraphics[width=1\columnwidth]{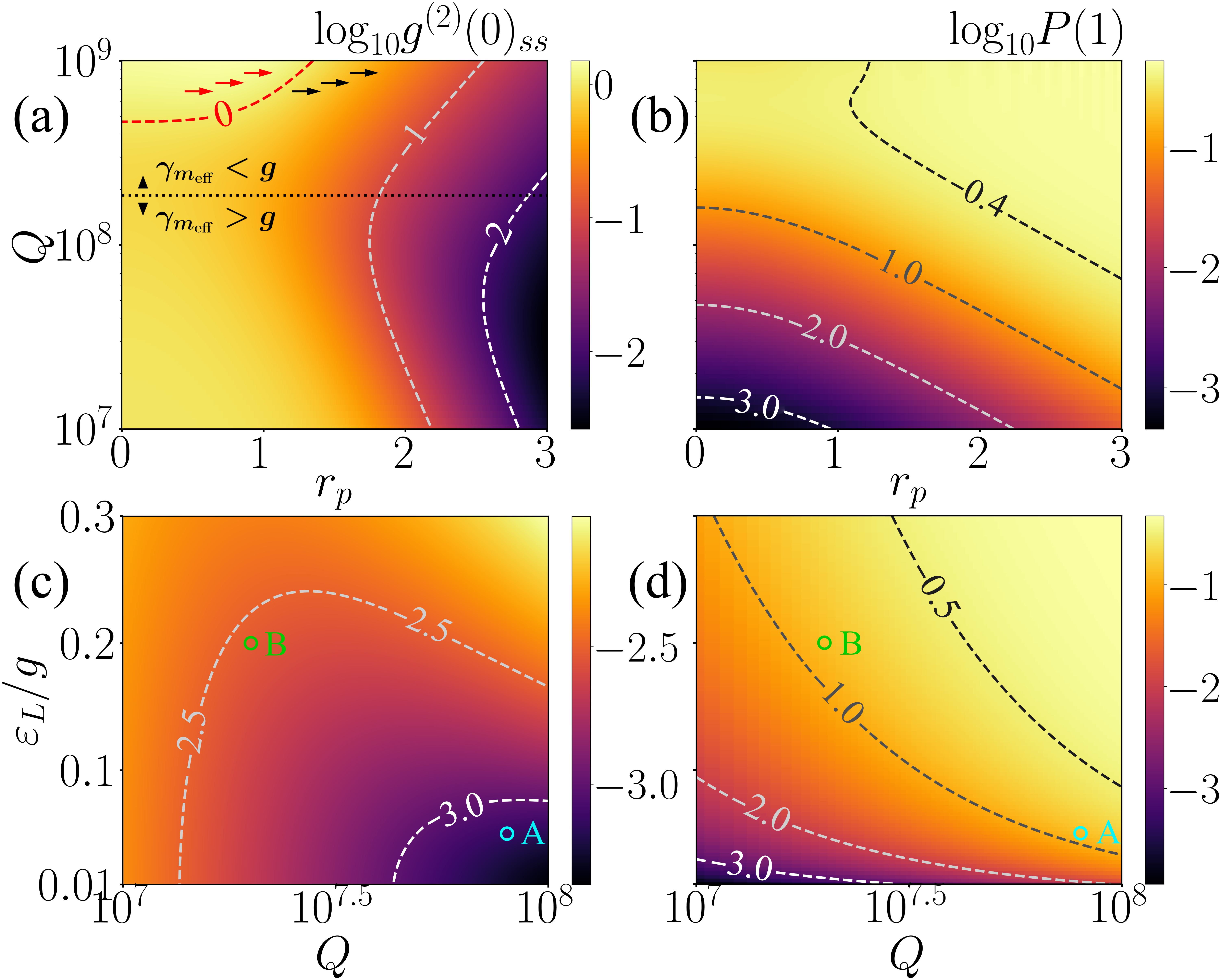}
	\caption{(a)-(b) Logarithmic second-order correlation function $g^{(2)}(0)_{ss}$ (a) and single-phonon probability $P(1)$ (b) versus the mechanical oscillator's $Q$ factor and the squeezing parameter $r_p$ for a fixed mechanical driving strength $\varepsilon_L=0.2g$. The horizontal dotted line corresponds to $\gamma_{m_\mathrm{{eff}}}=g$, which  divides the strong-coupling regime (upper) and the weak-coupling regime (lower) on both sides. (c)-(d) Logarithmic $g^{(2)}(0)_{ss}$ (c) and $P(1)$ (d) versus $Q$ and $\varepsilon_L$ for $r_p=3$. Point A illustrates a feasible point for ultra-strong PB $g^{(2)}(0)_{ss}<10^{-3}$ accompanied with an applicable single-phonon probability $P(1)>0.1$, with $Q=8\times10^7$ and $\varepsilon_L=0.05g$; Point B exemplifies a point with low $Q=2\times10^7$ and  $\varepsilon_L=0.2g$, corresponding to $g^{(2)}(0)_{ss}<10^{-2.5}$ and $P(1)>0.1$. The common parameters used here are the same as in Fig.~\ref{fig5}(a).}
	\label{fig6}
\end{figure}

We now proceed to show how to enhance the PB emerging in the weakly coupled system by modulating MPA. Specifically, we plot in Fig.~\ref{fig6}(a) the logarithmic second-order correlation function $g^{(2)}(0)_{ss}$ versus the squeezing parameter $r_p$ as well as the $Q$ factor  of the mechanical oscillator. As seen, for a fixed value of $Q$, the phonon antibunching effect can be gradually strengthened with increasing $r_p$. In particular, when $r_p\rightarrow3$, $g^{(2)}(0)_{ss}<10^{-2}$ is achievable even for $Q$ as low as $10^7$ (corresponding to $\gamma_{m_\mathrm{{eff}}}/g\approx20$). Note that here, in the region where strong PB occurs, the single-phonon probability can not be very high due to a low ratio $\varepsilon_L'/\gamma_{m_\mathrm{{eff}}}$. This is illustrated in Fig.~\ref{fig6}(b), where we see that the single-phonon probability $P(1)$ is lower than 0.1 for $r_p=3$ and $Q=10^7$, which can be increased to exceed 0.1 with increasing $Q$.
In addition, one sees in Fig.~\ref{fig6}(a) that for larger $Q$, e.g., $Q=10^8$, the reduction of $g^{(2)}(0)_{ss}$ with increasing $r_p\to3$ slows down gradually, in contrast to the case with smaller $Q$. 
This is because for $Q=10^8$, the effective driving strength $\varepsilon_L'$ enhanced by the strong MPA exceeds the mechanical decay rate $\gamma_{m_\mathrm{{eff}}}$, thus weakening the effect of PB in accordance with our previous analysis. In this case (with a relatively high-$Q$ mechanical oscillator), it is desirable to apply a weaker bare drive $\varepsilon_L$ to ensure strong PB being triggered. To this end, we plot in Fig.~\ref{fig6}(c) the logarithmic correlation $g^{(2)}(0)_{ss}$ as functions of the mechanical $Q$ factor and the driving strength $\varepsilon_L$ for a fixed squeezing parameter $r_p=3$. One clearly sees that strong-PB region with $g^{(2)}(0)_{ss}<10^{-3}$ emerges around $Q=10^8$ when applying a weak drive with strength $\varepsilon_L<0.1g$.  Correspondingly, the single-phonon probability $P(1)>0.1$ can be achieved in the region of $g^{(2)}(0)_{ss}<10^{-3}$, as we exemplify with point A in Figs.~\ref{fig6}(c) and \ref{fig6}(d). Also, point B in both plots exemplifies a typical case where both strong PB and a large single-phonon probability could coexist in our weak-coupling system with a low-$Q$ oscillator. {  Note that the performance of PB in the general hybrid system with single-phonon spin-mechanical coupling is relatively weak due to the limitation on the strength of MPA (see Appendix \ref{appenH} for a detailed discussion). Thus, our two-phonon-based scheme may provide an attractive platform to access PB with better performance and large-scale tunability.}

It is also interesting to examine the effect of the adjustable MPA on the phonon statistics when the system is originally in the strong-coupling regime. As shown in Fig.~\ref{fig6}(a), we observe a small region of phonon bunching (or, super-Poissonian phonon statistics) characterized by $g^{(2)}(0)_{ss}>1$, which appears for $\gamma_{m_\mathrm{{eff}}}<g$ and a small $r_p$. By gradually enhancing the mechanical amplification, phonon antibunching with $g^{(2)}(0)_{ss}<1$ emerges, and finally strong PB with $g^{(2)}(0)_{ss}<0.1$ is achieved. This reveals that our approach provides a flexible tunability of the phonon statistics, and thus could enable a number of applications in the quantum control of phonon as well as phonon-based quantum networks.

\subsection{Effects of spin dephasing and thermal phonons}
\begin{figure}
	\includegraphics[width=1\columnwidth]{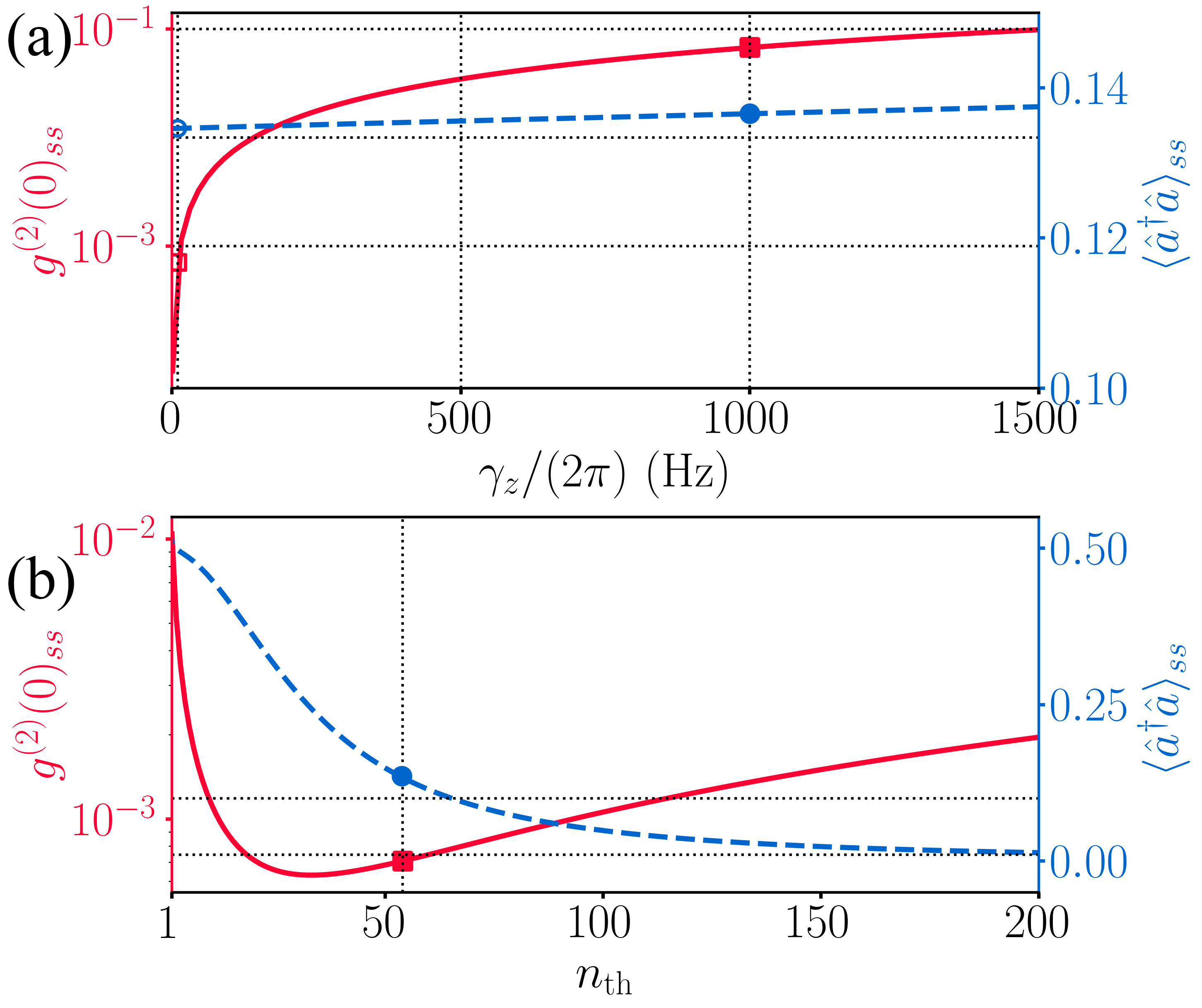}
	\caption{Steady-state second-order correlation function $g^{(2)}(0)_{ss}$ and mean phonon number $\langle \hat{a}^\dag\hat{a}\rangle_{ss}$ versus the qubit's dephasing rate $\gamma_z$ (a) and the mean thermal phonon number $n_{\mathrm{th}}$ (b). The hollow and solid markers in (a) correspond to $\gamma_z/(2\pi)=10$ Hz ($T_{2}\approx16$ ms) and $\gamma_z/(2\pi)=1$ kHz ($T_{2}^{\ast}\approx160$ $\mu$s), showing results of a future demonstration and within reach of current technology, respectively. The solid markers in (b) correspond to $n_{\mathrm{th}}=54$. The horizontal dotted lines in (b) denote $\langle \hat{a}^\dag\hat{a}\rangle_{ss}=0.1$ and 0.01.} The common parameters used here are the same as those for point A in Fig.~\ref{fig6}(c).
	\label{fig7}
\end{figure}

In the above analysis we have taken a fixed spin dephasing rate $\gamma_z=2\pi\times10$ Hz, corresponding to a dephasing time $T_{2}\approx16$ ms. This actually gives a future demonstration based on an assumption that the coherence properties of the shallow NV center we are considering can be greatly improved.  To further  evaluate the effect of varying $\gamma_z$ on the PB and also show the PB performance within the extent of current technology, we plot in Fig.~\ref{fig7}(a) the steady-state second-order correlation function $g^{(2)}(0)_{ss}$ and mean phonon number $\langle \hat{a}^\dag\hat{a}\rangle_{ss}$ versus $\gamma_z/(2\pi)\in\left[1,1500\right]$ Hz. The hollow markers locate $\gamma_z=2\pi\times10$ Hz ($T_{2}\approx16$ ms), which reproduce the results obtained by point A in Figs.~\ref{fig6}(c) and \ref{fig6}(d). With increasing $\gamma_z$ the value of $g^{(2)}(0)_{ss}$ increases sharply at first and then  slows down, while the value of $\langle \hat{a}^\dag\hat{a}\rangle_{ss}$ is nearly independent of varying $\gamma_z$. For $\gamma_z/(2\pi)=1$ kHz or $T_{2}^{\ast}\approx160$ $\mu$s, which are parameters accessible to current technology \cite{ishikawa2012optical}, PB with $g^{(2)}(0)_{ss}<10^{-1}$ and $\langle \hat{a}^\dag\hat{a}\rangle_{ss}>0.1$ can be observed (see solid markers in Fig.~\ref{fig7}(a)). 

Fig.~\ref{fig7}(b) shows the effects of thermal phonons on the performance of the enhanced PB. It is clear that increasing the mean thermal phonon number $n_{\mathrm{th}}$ does not destroy the PB effect significantly, as $g^{(2)}(0)_{ss}$ can still stay far below $10^{-2}$ when $n_{\mathrm{th}}$ increases to 200. By contrast, a large $n_{\mathrm{th}}$ has a greater effect on the mean phonon number in the mechanical oscillator, reducing $\langle \hat{a}^{\dagger} \hat{a} \rangle_{ss}$ to be about $0.013$ for $n_{\mathrm{th}}=200$. This suggests that the present protocol could yield better performance by operating at cryogenic temperatures.

\section{Detection of the PB effect}\label{sec4}
Unlike the photon blockade effects that can be measured directly via mature experimental technologies of photon correlation detection \cite{PhysRevLett.118.133604,PhysRevLett.106.243601,prasad2020correlating}, direct measurement of PB by measuring the second-order correlation function of phonons remains an experimental challenge. Nonetheless, it is feasible to measure the phonon correlations indirectly by, e.g., converting the mechanical signals into optical signals through auxiliary optomechanical couplings \cite{PhysRevLett.110.193602,cohen2015phonon,PhysRevA.94.063853,PhysRevA.98.013821}. A recent experiment demonstrated the measurement of second-order phonon correlation function in an optomechanical system by detecting the correlation of the emitted photons from the
optical cavity \cite{cohen2015phonon}. In addition, there have been numerous theoretical proposals that make it possible to implement the detection of PB. For example, it is demonstrated in Ref.~\cite{PhysRevA.82.032101} that PB could be measured by examining the power spectrum of the induced electromotive force between two ends of the mechanical oscillator. Also, PB could be accurately detected by measuring the statistics of photon in a superconducting microwave resonator which is resonantly coupled to the mechanical oscillator \cite{PhysRevB.84.054503}. This approach utilizes a perfect match between the phonon dynamics and the photon statistics, thus shifting the required measurement to the treatment of microwave photon detection. Moreover, a similar scheme exploiting an ancillary optical cavity to indirectly detect PB has been proposed \cite{PhysRevA.94.063853}.

In contrast to introducing additional optical or microwave cavity field, the qubit component involved in the primary system can also be used to detect PB. As demonstrated in Ref.~\cite{PhysRevA.93.063861}, besides using the charge qubit to produce PB states, the excited state probability of the qubit itself acts as an indication of PB. Note that this approach is well-suited for our scheme as well since the PB we predicted originates from the enhanced two-phonon nonlinearity between the mechanical resonator and the spin qubit. Following a similar way as in Ref.~\cite{PhysRevA.93.063861}, we next demonstrate how to qualitatively detect PB in our hybrid system by measuring the spin qubit. 

\begin{figure}
	\includegraphics[width=1\columnwidth]{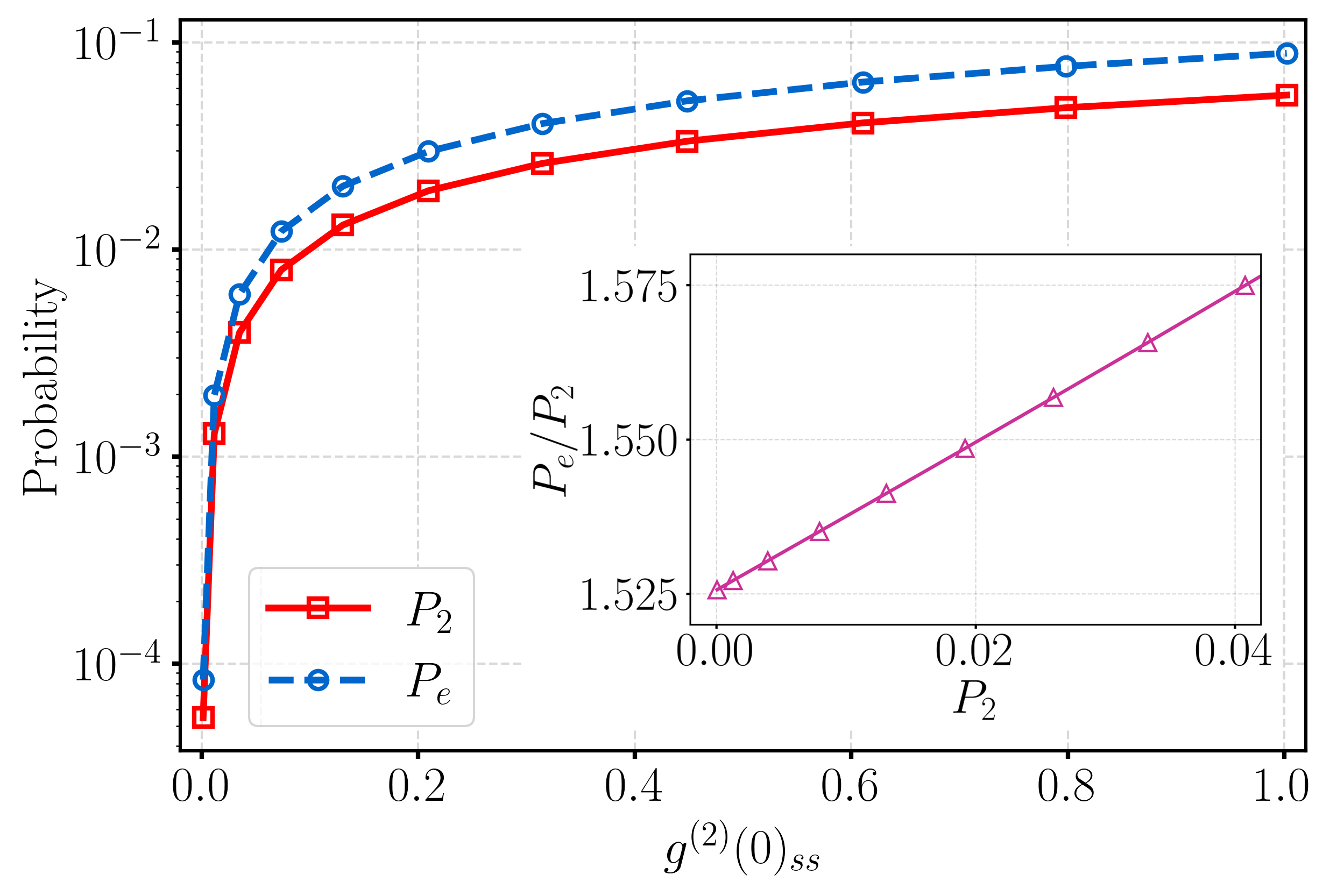}
	\caption{Probabilities $P_{2}$ and $P_{e}$ versus second-order correlation function $g^{(2)}(0)_{ss}$. The inset shows a ratio $P_{e}/P_{2}$, defined as detection sensitivity, versus $P_{2}$. The common parameters used here are the same as those for point A in Fig.~\ref{fig6}(c).}
	\label{fig8}
\end{figure}

The basic idea of the  detection relies on the fact that the probability of the second phonon is greatly suppressed when strong single PB occurs. On the contrary, an imperfect single PB takes in additional phonons, making the probability of the second phonon detectable. For convenience, we denote the probabilities of the mechanical oscillator in the Fock state $|2\rangle_{s}$ and the qubit in the excited state as $P_{2}$ and $P_{e}$, respectively. In the weak-driving regime, we have $P_2\approx |C_{2d}|^2$ and $P_e\approx |C_{0e}|^2$. Also, we see in Eq.~(\ref{c6}) that $C_{2d}$ is proportional to $C_{0e}$. This reveals that we are able to estimate the probability of the second phonon $P_2$ by measuring $P_e$. To verify this, we plot in Fig.~\ref{fig8} $P_{2}$ and $P_{e}$ versus $g^{(2)}(0)_{ss}$, by adjusting the driving strength $\varepsilon_{L}$. It is shown that the two probabilities are both very small ($<10^{-4}$) when $g^{(2)}(0)_{ss}\rightarrow0$, while they rise up rapidly with increasing $g^{(2)}(0)_{ss}$. Therefore, the observation that the qubit is in the excited state can be a signature of imperfect single PB. The larger the excited-state probability of the qubit being measured, the worse the effect of PB triggered in the system. In addition, we introduce the detection sensitivity $P_{e}/P_{2}$ and estimate its value in the inset of Fig.~\ref{fig8}. As seen, $P_{e}/P_{2}>1.5$ is always achievable, which increases the possibility of the detection in the case of low probabilities of $P_{2}$.

\section{Conclusion}\label{sec5}
We have presented an experimental method to induce strong PB in a weakly-coupled hybrid system, yielding simultaneously a large mean phonon number from the mechanical resonator. We show that the particular geometric arrangement of nanomagnets in our hybrid setup enables a quadratic interaction between the NV spin and the resonator, whose coupling can be significantly enhanced by modulating the mechanical parametric drive. This allows us to improve the performance of PB, making ultra-strong PB, featured by extremely small second-order correlation functions, and a large mean phonon number accessible even when the hybrid system is originally in the weak-coupling regime. Since the proposal studied here demonstrates the realization of enhanced PB in a realistic setup with experimentally accessible parameters, we hope that our work could provide a feasible and powerful tool  for exploring phonon statistics and practical applications such as single-phonon source or other single-phonon quantum devices.

\section{Acknowledgements}
We would like to thank Yuan Zhou for helpful discussions. This work was supported by National Natural Science Foundation of China (NSFC) (11675046), Program for Innovation Research of Science in Harbin Institute of Technology (A201412), and Postdoctoral Scientific Research Developmental Fund of Heilongjiang Province (LBH-Q15060).

\appendix
{ \section{Experimental realizations of device fabrication and arrangement}\label{appenA}
 From an application standpoint, all components required for the proposed setup could be fabricated individually with modern nanofabrication techniques, though they have not been combined in a single experiment. Specifically, diamond cantilevers embedded with NV centers with thickness less than 200 nm have been used for achieving strong NV-strain coupling \cite{PhysRevApplied.5.034010}. Notably, diamond nanocantilevers of similar dimensions to that used in our simulations have been fabricated using an angled-etching technique \cite{doi:10.1063/1.4937625,burek2012free}. Also, the introduction of NV centers into diamond has been achieved through various techniques, such as nitrogen ion implantation for bulk diamond samples \cite{maletinsky2012robust,ovartchaiyapong2014dynamic} and nitrogen doping during chemical vapor deposition growth of thin (e.g., 5 nm) diamond films \cite{doi:10.1063/1.4748280,ohashi2013negatively}. As for the Dy nanomagnet, it could be fabricated in experiments via a lift-off process using electron-beam deposition as demonstrated in Ref.~\cite{doi:10.1063/1.3673910}, where cylinder-like Dy tips with high magnetic field gradients have been fabricated for application of magnetic resonance force microscopy. To magnetically couple the NV spin and the diamond cantilever's position, the two nanomagnets should be nano-positioned symmetrically on both sides of the cantilever \cite{kolkowitz2012coherent}, and the position optimized to generate the gradient along the NV axis. Having settled the position the nanomagnets should be maintained at a fixed height (with respect to the cantilever's upper and lower surfaces) during the experiment, while the cantilever driven at its resonance frequency using, e.g., a piezoelectric module \cite{arcizet2011single,kolkowitz2012coherent}.
}

\begin{figure}
	\includegraphics[width=1\columnwidth]{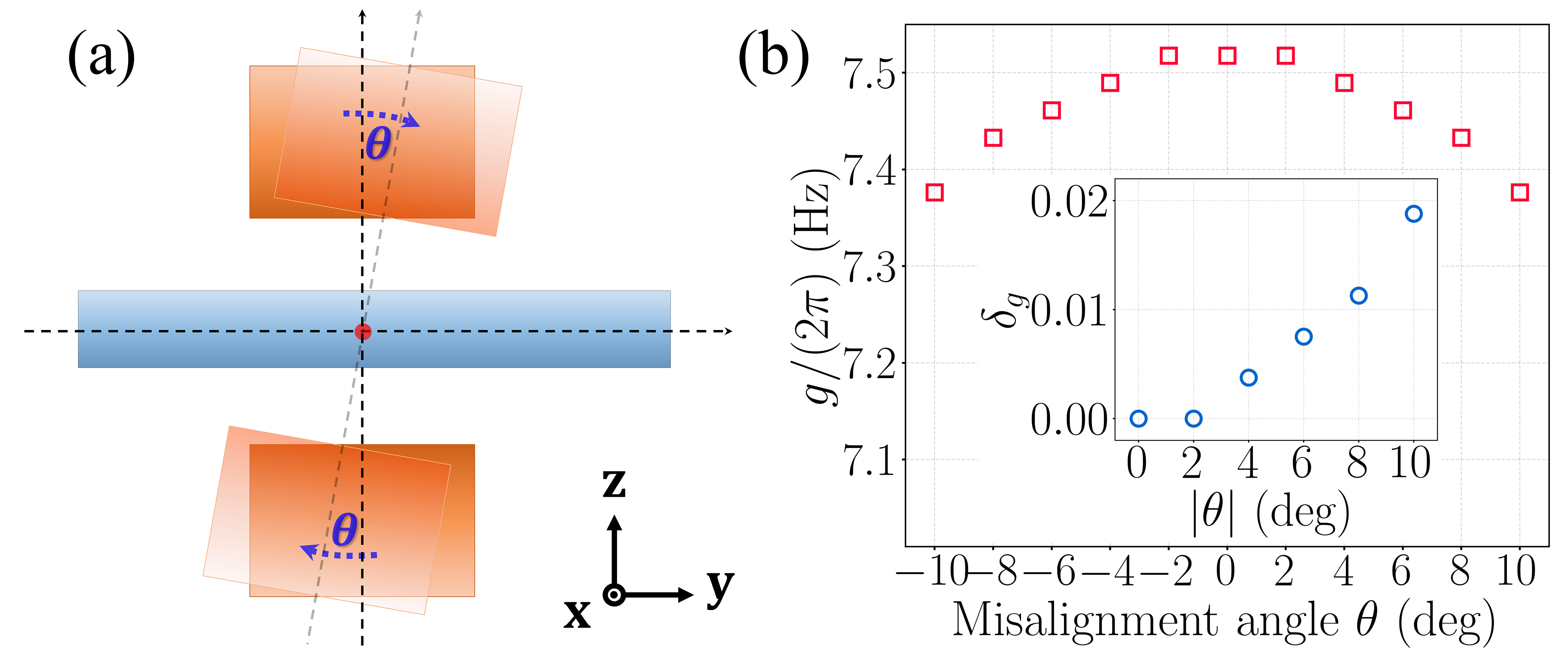}
	\caption{{ (a) Schematic diagram showing field orientation misalignment due to imperfectly positioned magnets. Here we assume that the two magnets are fixed together and misaligned by an angle $\theta$ with respect to the $z$ axis. (b) The two-phonon coupling rate $g$ as a function of $\theta$. Positive and negative $\theta$ indicate clockwise and counterclockwise tilting of magnets, respectively. The insets shows the relative error $\delta_g$ versus $\theta$. The common parameters are the same as in Fig.~\ref{fig2}(a).}}
	\label{figA0}
\end{figure}

{ \section{Effect of field orientation misalignment}\label{appenB}
As shown in Fig.~\ref{figA0}(a), we consider the case where the two nanomagnets are misaligned by a small angle $\theta$ with respect to the $z$ axis, which could be caused by imperfect operations in the experiment. In this case, the magnetic field generated by the tilted magnets has an extremum at the position of the NV center, thus providing only second-order coupling to the mechanical mode. In the absence of the misalignment, the magnets generate a field that has null second derivatives along the $x$ and $y$ axes, thus giving a single second-order magnetic gradient along the $z$ axis, $G_z=\partial^{2} B_z / \partial z^{2}(0)$. However, for the case considered here the misalignment also introduces a second-order magnetic gradient along the $y$ axis, while it does not contribute to the spin-mechanical coupling due to the assumption that the mechanical mode oscillates only along the $z$ direction. Based on finite-element simulations we acquire the dependence of the gradient $G_z$ on the misalignment angle $\theta$ and then calculate the two-phonon coupling rate $g$ versus $\theta$, as depicted in Fig.~\ref{figA0}(b). We observe that the coupling $g$ is reduced slightly with increasing the tilting angle $\theta$, and a misalignment of $10^\circ$ can result in a relative error $\delta_g$ of less than $2\%$, with $\delta_g=(g_\theta-g_{\theta=0})/g_{\theta=0}$.
}

\section{Derivation of the two-phonon Jaynes-Cummings  Hamiltonian}\label{appenC}
\begin{figure}
	\includegraphics[width=1\columnwidth]{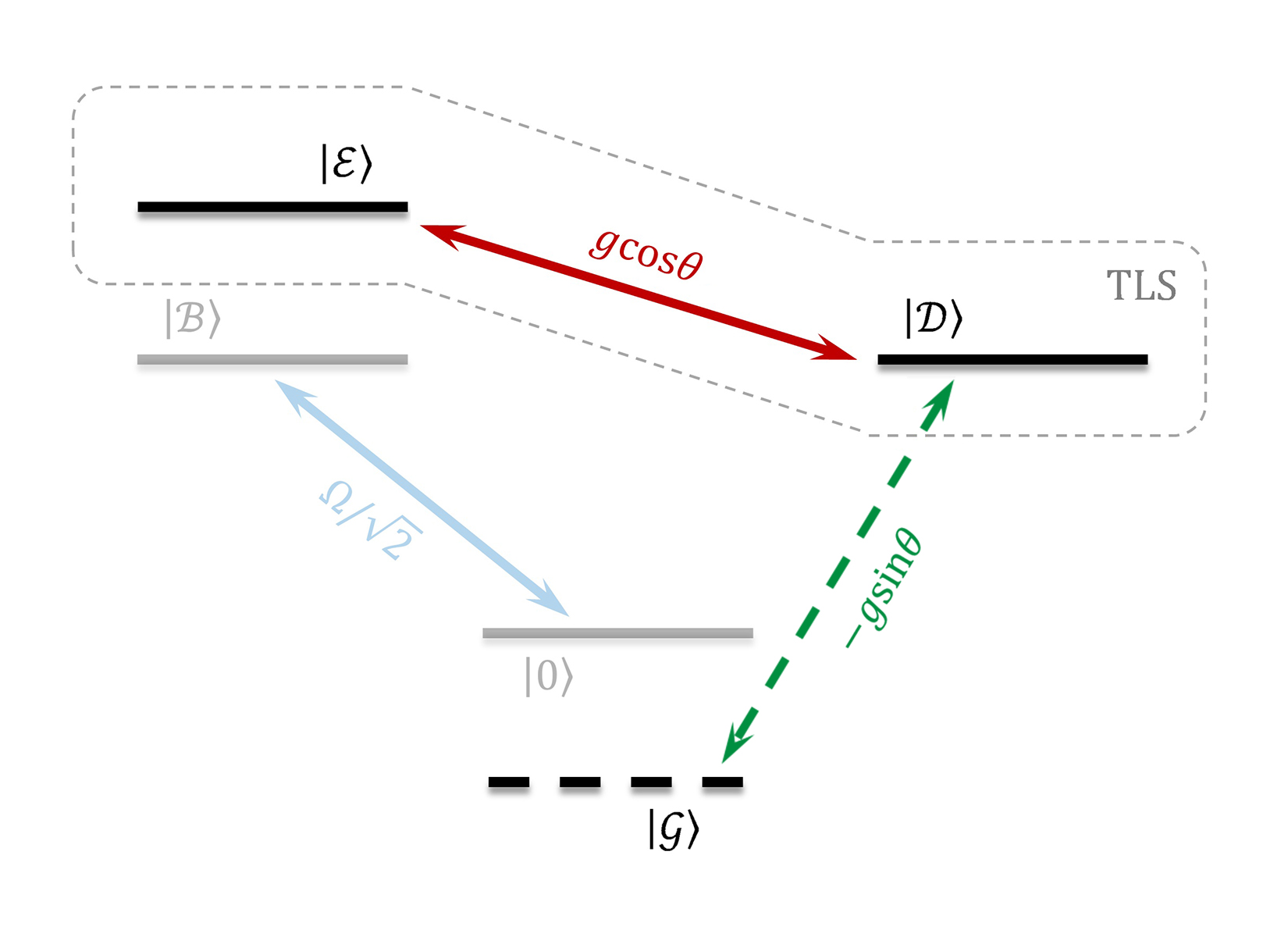}
	\caption{Level diagram of the dressed spin basis states. The states $|\mathcal{E}\rangle$ and $|\mathcal{D}\rangle$ construct an effective TLS where the effective coupling strength between states is approximated as $g$ (i.e., $\cos\theta\simeq1$) for $\Delta\gg\Omega$.}
	\label{figA1}
\end{figure}

To begin with, we recall the total Hamiltonian given by Eq.~(\ref{H_total})
\begin{equation}
\begin{split}
\hat{H}_{\mathrm{Total}}=&\hat{H}_{\mathrm{mec}}+\hat{H}_{\mathrm{NV}}+\hat{H}_{\mathrm{int}}\\
=&\omega_{m} \hat{a}^{\dagger} \hat{a}+\sum_{j=\pm}\Delta|j\rangle\langle j|+\frac{\Omega}{2}(|0\rangle\langle j|+| j\rangle\langle 0|)\\
&+g\left( \hat{a}^{\dagger}+\hat{a}\right)^2  \hat{S}_z+\Omega_{p} \cos \left(2 \omega_{p} t\right)\left(\hat{a}^{\dagger}+\hat{a}\right)^{2}
\\
&+\varepsilon_L\left(\hat{a}^{\dagger}e^{-i\omega_L t}+ \hat{a} e^{i\omega_L t}\right).
\end{split}
\end{equation}
Looking at the spin-only part, it is clear that $\hat{H}_{\mathrm{NV}}$ couples the state $|0\rangle$ to a bright superposition of excited states $|\mathcal{B}\rangle=(|+1\rangle+|-1\rangle) / \sqrt{2}$, whereas the dark superposition $|\mathcal{D}\rangle=(|+1\rangle-|-1\rangle)/\sqrt{2}$ remains decoupled. Therefore, the eigenbasis of
$\hat{H}_{\mathrm{NV}}$ is given by $|\mathcal{D}\rangle$ and two dressed states $|\mathcal{G}\rangle=\cos \theta|0\rangle-\sin \theta|\mathcal{B}\rangle,|\mathcal{E}\rangle=\cos \theta|\mathcal{B}\rangle+\sin \theta|0\rangle\}$, with $\tan (2 \theta)=\sqrt{2} \Omega / \Delta$. The corresponding eigenfrequencies are $\omega_d=\Delta$ and $\omega_{e(g)}=\left(\Delta \pm \sqrt{\Delta^{2}+2 \Omega^{2}}\right)/2$. The level diagram of the dressed spin states is illustrated in Fig.~\ref{figA1} . By working in the dressed-state basis $\{|\mathcal{D}\rangle,|\mathcal{G}\rangle,|\mathcal{E}\rangle\}$, $\hat{H}_{\mathrm{NV}}$ can be diagonalized, and the total Hamiltonian is rewritten as \cite{PhysRevLett.117.015502}
\begin{equation} \label{A2}
\begin{split}
\hat{H}_{\mathrm{Total}}=&\omega_{m} \hat{a}^{\dagger} \hat{a}+\omega_{eg}|\mathcal{E}\rangle\left\langle \mathcal{E}\left|+\omega_{dg}\right| \mathcal{D}\right\rangle\langle \mathcal{D}|\\
&+ g\left( \hat{a}^{\dagger}+\hat{a}\right)^2\left(\cos\theta|\mathcal{E}\rangle \langle \mathcal{D}|-\sin\theta|\mathcal{G}\rangle \langle \mathcal{D}|+\mathrm{H.c.} \right)\\
&+\Omega_{p} \cos \left(2 \omega_{p} t\right)\left(\hat{a}^{\dagger}+\hat{a}\right)^{2}
\\
&+\varepsilon_L\left(\hat{a}^{\dagger}e^{-i\omega_L t}+ \hat{a} e^{i\omega_L t}\right),
\end{split}
\end{equation}
where $\omega_{eg}=\sqrt{\Delta^{2}+2 \Omega^{2}}$ and $\omega_{dg}=\left( \Delta+\sqrt{\Delta^{2}+2 \Omega^{2}}\right)/2$. In the limit $\Delta\gg\Omega$, we have $\cos\theta\simeq1$, $\sin\theta\simeq0$, $\omega_{eg}\simeq\Delta+\Omega^2/\Delta$, $\omega_{dg}\simeq\Delta+\Omega^2/2\Delta$ and $|\mathcal{E}\rangle\simeq|\mathcal{B}\rangle$. This implies a particular case where $\omega_{dg}\gg\omega_{ed}\simeq\Omega^2/2\Delta$, thus allowing us to choose a new basis   $\left\lbrace |\mathcal{E}\rangle,|\mathcal{D}\rangle\right\rbrace $ constructing an effective two-level system (TLS, see Fig.~\ref{figA1}). In this case, Eq.~(\ref{A2}) is simplified as
\begin{equation} \label{A3}
\begin{split}
\hat{H}_{\mathrm{Total}}\simeq&\omega_{m} \hat{a}^{\dagger} \hat{a}+\omega_{ed}\hat{\sigma}_{+}\hat{\sigma}_{-}+ g\left( \hat{a}^{\dagger}+\hat{a}\right)^2\left(\hat{\sigma}_{+}+\hat{\sigma}_{-}\right)\\
&+\Omega_{p} \cos \left(2 \omega_{p} t\right)\left(\hat{a}^{\dagger}+\hat{a}\right)^{2}
\\
&+\varepsilon_L\left(\hat{a}^{\dagger}e^{-i\omega_L t}+ \hat{a} e^{i\omega_L t}\right).
\end{split}
\end{equation}
where $\hat{\sigma}_{+} \equiv|\mathcal{E}\rangle\langle \mathcal{D}|$ and $\hat{\sigma}_{-} \equiv|\mathcal{D}\rangle\langle \mathcal{E}|$ are the raising and lowering operators of the effective TLS, respectively. Further, by performing a unitary transformation $\hat{U}=e^{-i \hat{H}_{0} t}$ with $ \hat{H}_{0}=\omega_{p}\left(\hat{a}^{\dagger} \hat{a}+2|e\rangle\langle e|\right)$ for Eq.~(\ref{A3}) and considering $\omega_p\gg\left\lbrace \Omega_p,g \right\rbrace $ for the RWA, we end up with
\begin{equation}\label{A4}
\begin{split}
\hat{H}_{\mathrm{Total}}\simeq &\delta_{m} \hat{a}^{\dagger} \hat{a}+\delta_{ed}\hat{\sigma}_{+}\hat{\sigma}_{-}+ g\left( \hat{a}^{\dagger2}\hat{\sigma}_{-}+\hat{a}^{2}\hat{\sigma}_{+}\right)\\
&+\frac{\Omega_{p}}{2} \left(\hat{a}^{\dagger 2}+\hat{a}^{2}\right)+\varepsilon_L\left(\hat{a}^{\dagger}e^{-i\delta_L t}+ \hat{a} e^{i\delta_L t}\right).
\end{split}
\end{equation}
where $\delta_{m(L)}=\omega_{m(L)}-\omega_{p}$ and $\delta_{ed}=\omega_{ed}-2\omega_{p}$.

\section{Hybrid-system Hamiltonian in the squeezed frame}\label{appenD}
Considering the system Hamiltonian given in Eq.~(\ref{A4}), we perform a unitary squeezing transformation $\hat{U}_{s}=\exp \left[r_p\left(\hat{a}^{2}-\hat{a}^{\dagger 2}\right) / 2\right]$, where the squeezing parameter $r_p$ is defined via $r_{p}=(1 / 2) \operatorname{arctanh}\left(\Omega_{p} / \delta_{m}\right)$. Then, the system Hamiltonian transformed to the squeezed reference frame reads
\begin{equation}\label{B1}
\begin{split}
\hat{H}_{\mathrm{Total}}^{\mathrm{S}}= &\delta_{s} \hat{a}_{s}^{\dagger} \hat{a}_{s}+\delta_{ed}\hat{\sigma}_{+}\hat{\sigma}_{-}+ gU^2(\hat{a}_{s}^{\dagger2}\hat{\sigma}_{-}+\hat{a}_{s}^{2}\hat{\sigma}_{+})\\
&+g V^2 (\hat{a}_{s}^{\dagger2}\hat{\sigma}_{+}+\hat{a}_{s}^{2}\hat{\sigma}_{-})-gUV(\hat{a}_{s}^{\dagger} \hat{a}_{s}+ \hat{a}_{s} \hat{a}_{s}^{\dagger})\\
&\times(\hat{\sigma}_{+}+\hat{\sigma}_{-})+\varepsilon_LU\left(\hat{a}_{s}^{\dagger}e^{-i\delta_L t}+\hat{a}_{s}e^{i\delta_L t}\right)\\
&-\varepsilon_LV\left(\hat{a}_{s}^{\dagger}e^{i\delta_L t}+\hat{a}_{s}e^{\mathrm{-i}\delta_L t}\right),
\end{split}
\end{equation}
where $\delta_{s}=\delta_{m}/\cosh(2r_p)$ represents the squeezed-oscillator frequency, $U=\cosh(r_p)$ and $V=\sinh(r_p)$. { Note that in Eq.~(\ref{B1}) the bare mechanical mode $\hat{a}$ in the original lab frame has been transformed to a squeezed mode $\hat{a}_{s}$ in the squeezed frame.} Under the two-phonon resonance condition $\delta_{ed}=2\delta_s$, the two terms with coefficients $gV^2$ and $-gUV$ in Eq.~(\ref{B1}) are off-resonant and hence could be discarded via RWA, e.g., assuming a large detuning condition $\delta_{ed}\gg gUV$. Further, for a near-resonant mechanical drive $\delta_s\approx\delta_L$, the term with coefficient $-\varepsilon_LV$ is largely detuned under the weak-drive condition $\varepsilon_L\ll g$ (giving $\varepsilon_LV\ll2\delta_s$).
Therefore, we obtain the simplified system Hamiltonian in the squeezed frame, with an effective form 
\begin{equation}\label{B2}
\begin{split}
\hat{H}_{\mathrm{Total}}^{\mathrm{S}}\simeq &\delta_{s} \hat{a}_{s}^{\dagger} \hat{a}_{s}+\delta_{ed}\hat{\sigma}_{+}\hat{\sigma}_{-} +g_{\mathrm{eff}}(\hat{a}_{s}^{\dagger2}\hat{\sigma}_{-}+\hat{a}_{s}^{2}\hat{\sigma}_{+})\\
&+\varepsilon_L'\left(\hat{a}_{s}^{\dagger}e^{-i\delta_L t}+\hat{a}_{s}e^{i\delta_L t}\right),
\end{split}
\end{equation}
where we have defined $g_{\mathrm{eff}}=gU^2$ and $\varepsilon_L'=\varepsilon_LU$ to lighten the
notation. A direct demonstration of the validity of Eq.~(\ref{B2}) is given in Fig.~\ref{fig5}(h),  where the evolution of states obtained using the effective Hamiltonian (\ref{B2}) agrees well with that obtained using the exact Hamiltonian (\ref{B1}).

{ 
\section{Engineering squeezed reservoir for suppressing mechanical noise}\label{appenF}
As we mentioned in the main text, introducing MPA to enhance the spin-mechanical coupling also enhances the coupling between the mechanical mode and its bath environment, resulting in a significantly amplified mechanical noise. A promising strategy against this problem is to use the squeezed-vacuum-reservoir technique, which  has been studied extensively in recent reports \cite{PhysRevLett.114.093602,PhysRevLett.120.093601,PhysRevLett.120.093602,PhysRevA.99.023833,https://doi.org/10.1002/andp.201900220,PhysRevA.100.012339,PhysRevA.100.062501,PhysRevA.101.053826,PhysRevA.102.032601,PhysRevLett.126.023602}. In what follows, we first sketch out the physical mechanism of the strategy and then show how to integrate it into our proposed device by engineering a microwave cavity-optomechanical system. 

Squeezing the mechanical mode induces an increase in system-reservoir coupling with a rate proportional to exponential squeezing parameter $r_p$. In other words, lab-frame vacuum noise becomes squeezed in the squeezed frame, and introduces dissipative dynamics that have a similar form to thermal dissipation. By  introducing the squeezed-vacuum-reservoir technique as proposed in Ref.~\cite{PhysRevLett.114.093602}, the squeezing-enhanced mechanical noise in our scheme could be effectively suppressed. The technique depends on introducing an auxiliary, broadband squeezed-vacuum field to drive the cavity, which is phase matched with the parametric amplification that squeezes the cavity mode. This ensures that the squeezed cavity mode is equivalently coupled to a thermal vacuum reservoir, thereby allowing to describe the system dynamics with a simplified master equation in the standard Lindblad form (see Refs.~\cite{PhysRevLett.114.093602,PhysRevLett.120.093601,PhysRevA.99.023833} for more technical details). 

For our scheme, the above auxiliary-reservoir technique could be implemented in practice by integrating an appropriate microwave optomechanical system into the present device (as depicted in Fig.~\ref{fig1}). Here we schematically illustrate a microwave circuit diagram that includes only additional components to Fig.~\ref{fig1}, as shown in Fig.~\ref{figA4}. The circuit consists of a vibrating capacitor coupled to a superconducting microwave resonator terminated by a superconducting quantum interference device (SQUID). The capacitor consists of two electrodes, one of which is coated on the cantilever while another placed	close and parallel to the cantilever (see the inset in Fig.~\ref{figA4}), which couples electrical energy to mechanical motion. Note that such an arrangement is similar to that for activating the process of MPA, and has been demonstrated in some experiments \cite{1438421,LUO2006139}. The microwave resonator  connecting a SQUID (equivalent to a Josephson parametric amplifier) provides squeezed microwave inputs to the mechanical mode; For sufficiently large bandwidth of the microwave squeezing, the microwave resonator could be regarded as an effective squeezed-vacuum reservoir of the mechanical mode. Experimentally, squeezing bandwidths up to several hundreds of MHz \cite{doi:10.1063/1.4886408,doi:10.1063/1.4939148,doi:10.1063/5.0035945}, or even $\sim$ GHz \cite{PhysRevX.10.021021}, have been reported, which is sufficient to fulfill the large-bandwidth requirement of the reservoir. The squeezing parameter and the reference phase of the effective reservoir, as main controlled parameters for the auxiliary-reservoir technique, can be controlled by the amplitude and phase of the pump tone used to modulate the magnetic flux ($\Phi$) through the SQUID.

\begin{figure}
	\includegraphics[width=1\columnwidth]{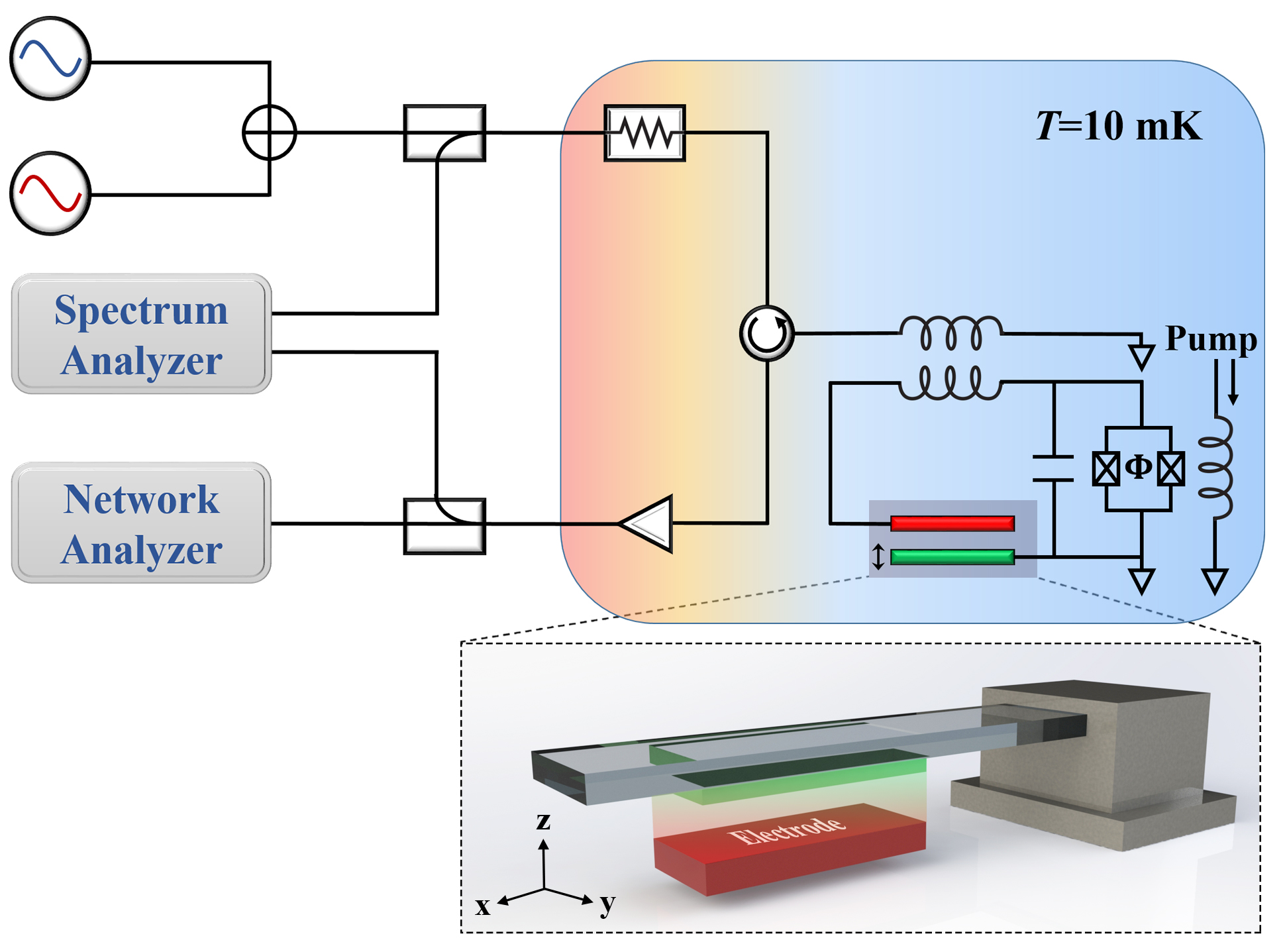}
	\caption{{ Simplified schematic diagram showing a microwave circuit for squeezed reservoir engineering and for initial state preparation of  mechanical squeezed vacuum. The microwave cavity-optomechanical system is a LC circuit composed of a spiral inductor and a vibrating capacitor. The capacitor shown in the inset consists of two electrodes, one coated on the cantilever while the other placed close and parallel to the cantilever. A flux-pumped SQUID connected to the microwave resonator produces squeezed microwave field for the mechanical oscillator. Two pump tones are filtered and attenuated before reaching the system mounted in a dilution refrigerator (with a temperature $T=10$ mK). Signals from the device are amplified and then measured with a spectrum analyzer or network analyzer.}}
	\label{figA4}
\end{figure}

In addition to suppress the squeezing-enhanced mechanical noise, the auxiliary microwave resonator can be conveniently used for initial state preparation of mechanical squeezed vacuum. The basic idea is based on the dissipative squeezing method \cite{PhysRevA.88.063833}, which has been experimentally demonstrated effective in cooling the mechanical resonator (into a squeezed state) \cite{wollman2015quantum,PhysRevLett.115.243601,PhysRevX.5.041037,toth2017dissipative,PhysRevLett.123.183603}.
As schematically shown in Fig.~\ref{figA4}, two microwave pumps of different frequency $\omega_{+,-}$ are filtered at room temperature and attenuated at a low temperature of 10 mK, and then inductively coupled to the microwave resonator. By standard linearization with a displacement transformation $\hat{c}=\bar{c}_{+} e^{-i \omega_{+} t}+\bar{c}_{-} e^{-i \omega_{-} t}+\hat{d}$, where $\hat{c}$ is the annihilation operator of the cavity field and $\bar{c}_{+,-}$ are the cavity field amplitude due to the two pumps, the resulting optomechanical Hamiltonian, in the interaction picture, can be written as
\begin{equation}\label{H_lin}
\begin{aligned}
\hat{H}_{\mathrm{lin}}=&-\hat{d}^{\dagger}\left(G_{+} \hat{a}^{\dagger}+G_{-} \hat{a}\right)- \hat{d}^{\dagger}(G_{+} \hat{a} e^{-2 i \delta_m t}\\
&+G_{-} \hat{a}^{\dagger} e^{2 i \delta_m t})+\mathrm{H.c.},
\end{aligned}
\end{equation}
where $G_{+,-}$ are the driven-enhanced optomechanical coupling rates. Further, assuming $G_{+}<G_{-}$ and by rewriting the Hamiltonian (\ref{H_lin}) using a Bogoliubov transformation, one finds that the mode $\hat{d}$ is coupled directly to an effective mode $\hat{\beta}=\cosh (r) \hat{a}+\sinh (r) \hat{a}^{\dagger}$ with $r=\tanh^{-1}(G_{+}/G_{-})$, via a beam-splitter
Hamiltonian
\begin{equation}\label{H_cool}
\hat{H}_{\mathrm{lin}}\simeq-\mathcal{G} \hat{d}^{\dagger} \hat{\beta}+\text {H.c.},
\end{equation}
where $\mathcal{G}=\sqrt{G_{-}^{2}-G_{+}^{2}}$. In fact, the Bogoliubov mode $\hat{\beta}$ can be expressed as a transformation of the mode $\hat{a}$ by the squeezing operator $\hat{S}(r)=\exp \left[r\left( \hat{a}^{2}-\hat{a}^{\dagger 2}\right)/2\right]$, i.e., $\hat{\beta}=\hat{S}(r)\hat{a}\hat{S}^{\dagger}(r)$. The Hamiltonian (\ref{H_cool}) indicates sideband cooling of the Bogoliubov mode to the vacuum state, which in the lab frame corresponds to cooling the mechanical mode towards a squeezed vacuum state.
}

\section{Approximate analytical expression of the second-order correlation function}\label{appenG}
We have obtained the equations of motion for the probability amplitudes of the four basis states, which are given by
\begin{align}
\dot{C}_{1 d}&=-i \varepsilon_{L}' C_{0d}-i\left(\delta-i\frac{\gamma_{m_\mathrm{{eff}}}}{2}\right) C_{1 d}-i\sqrt{2} \varepsilon_{L}' C_{2 d},\\
\dot{C}_{2 d}&=-i\sqrt{2} \varepsilon_{L}' C_{1d}-i\left(2\delta-i\gamma_{m_\mathrm{{eff}}}\right) C_{2d}-i\sqrt{2} g_{\mathrm{eff}} C_{0 e}, \\
\dot{C}_{0 e}&=-i\sqrt{2} g_{\mathrm{eff}} C_{2d}-i\left(2 \delta-i\frac{\gamma_{\mathrm{z}}}{2}\right)  C_{0 e}.
\end{align}
Upon setting $\dot{C}_{i j}= 0$ ($i \in \{0, 1, 2\}$ and $j \in \{ d, e\}$), the steady-state solution for each coefficient can be found by solving the following equations
\begin{align}
\label{c4} 0&=-i \varepsilon_{L}' C_{0d}-i\left(\delta-i\frac{\gamma_{m_\mathrm{{eff}}}}{2}\right) C_{1 d}-i\sqrt{2} \varepsilon_{L}' C_{2 d},\\
\label{c5} 0&=-i\sqrt{2} \varepsilon_{L}' C_{1d}-i\left(2\delta-i\gamma_{m_\mathrm{{eff}}}\right) C_{2d}-i\sqrt{2} g_{\mathrm{eff}} C_{0 e}, \\
\label{c6} 0&=-i\sqrt{2} g_{\mathrm{eff}} C_{2d}-i\left(2 \delta-i\frac{\gamma_{\mathrm{z}}}{2}\right)  C_{0 e}.
\end{align}
Combining Eq.~(\ref{c5}) with Eq.~(\ref{c6}) we obtain the relation between $C_{1d}$ and $C_{2d}$,
\begin{equation} \label{c7}
C_{2 d}=\frac{\sqrt{2}\varepsilon_{L}'\left(2\delta- i \gamma_\mathrm{z}/2\right) }{2g_{\mathrm{eff}}^{2}-\left(2\delta-i \gamma_{m_\mathrm{{eff}}}\right)\left(2\delta-i \gamma_{\mathrm{z}}/2\right) } C_{1 d}.
\end{equation}

Considering a sufficiently-weak mechanical drive allows us to assume $\left|C_{2 d}\right|^{2} \ll \min \left\lbrace \left|C_{0 d}\right|^{2},\left|C_{1d}\right|^{2}\right\rbrace $ and $\left|C_{0 d}\right|^{2}+\left|C_{1 d}\right|^{2} \approx 1$. Then, by neglecting $C_{2d}$ in Eq.~(\ref{c4}), we have 
\begin{equation} \label{c8}
|C_{1d}|^2=\frac{\varepsilon_{L}^{\prime 2}}{\delta^2+\varepsilon_{L}^{\prime 2} +(\gamma_{m_\mathrm{{eff}}}/2)^2}.
\end{equation}
Substituting Eq.~(\ref{c8}) into Eq.~(\ref{c7}), we can obtain 
\begin{equation} \label{c9}
|C_{2 d}|^2=\frac{2\varepsilon_{L}^{\prime 4}\left[ 4\delta^2+(\gamma_{\mathrm{z}}/2)^2\right] }{\left( 4\delta^4+\delta^2\zeta+\Xi^2\right)\left[ 4(\delta^2+\varepsilon_{L}^{\prime 2})+\gamma_{m_\mathrm{{eff}}}^2\right] }.
\end{equation}
where $\zeta=(\gamma_{\mathrm{z}}/2)^2+\gamma_{m_\mathrm{{eff}}}^2-4g_{\mathrm{eff}}^2$ and $\Xi=g_{\mathrm{eff}}^2+\gamma_{\mathrm{z}}\gamma_{m_\mathrm{{eff}}}/4$.

For the state given in Eq.~(\ref{psi}), the steady-state, equal-time second-order correlation function can be written, according to its definition in Eq.~(\ref{g2}), as 
\begin{equation}
\begin{split}
g^{(2)}(0)_{ss}&=\frac{2\left|C_{2d}\right|^{2}}{\left(\left|C_{1d}\right|^{2}+2\left|C_{2d}\right|^{2}\right)^{2}}\simeq\frac{2\left|C_{2d}\right|^{2}}{\left|C_{1d}\right|^{4}}\\
&\simeq\frac{\left[4\delta^2+(\gamma_{\mathrm{z}}/2)^2\right]\left[\delta^2+\varepsilon_{L}^{\prime 2} +(\gamma_{m_\mathrm{{eff}}}/2)^2\right]}{4\delta^4+\delta^2\zeta+\Xi^2  }.
\end{split}
\end{equation}

{ 
\section{PB Performance in the system with single-phonon coupling}\label{appenH}
In the main text, we have focused on the enhanced performance of PB in our system with a two-phonon nonlinearity. Here, as a contrast, we turn to a more general system with single-phonon spin-mechanical coupling and  examine the performance of PB in such a system in the presence of MPA. Following the normal procedures as given in Appendixes \ref{appenC} and \ref{appenD}, it is straightforward to obtain the system Hamiltonian in the squeezed frame, which is given by
\begin{equation}\label{F1}
\begin{split}
\hat{H}_{\mathrm{Total}}^{\mathrm{S}'}=&\delta_{s} \hat{a}_{s}^{\dagger} \hat{a}_{s}+\delta_{ed}'\hat{\sigma}_{+}\hat{\sigma}_{-}+g_0U(\hat{a}_{s}^{\dagger}\hat{\sigma}_{-}+\hat{a}_{s}\hat{\sigma}_{+})\\
&-g_0V(\hat{a}_{s}\hat{\sigma}_{-}+\hat{a}_{s}^{\dagger}\hat{\sigma}_{+})+\varepsilon_{L}U\left(\hat{a}_{s}^{\dagger}e^{-i\delta_L t}+\hat{a}_{s}e^{i\delta_L t}\right)\\
&-\varepsilon_LV\left(\hat{a}_{s}^{\dagger}e^{i\delta_L t}+\hat{a}_{s}e^{\mathrm{-i}\delta_L t}\right),
\end{split}
\end{equation}
where $\delta_{s}=\delta_{m}/\cosh(2r_p)$ with  $\delta_{m}=\omega_{m}-\omega_{p}$, $\delta_{ed}'=\omega_{ed}-\omega_{p}$, $\delta_{L}=\omega_{L}-\omega_{p}$, and $g_0 =\mu_{B} g_{e} z_{\mathrm{zpf}} G_{0}$ is the bare single-phonon coupling strength, with $G_{0}$ the first-order magnetic field gradient. State-of-the-art magnetic tips can provide a magnetic field gradient $G_{0}$ up to $10^6$ T/m \cite{Lee_2017}, giving a single-phonon coupling rate $g_0/2\pi\approx2.5$ kHz for the proposed diamond cantilever. Under the conditions of $\delta_s=\delta_{ed}\approx\delta_L$ and $\delta_s\gg g_{0}V,\varepsilon_LV$, the two terms with coefficients $gV$ and $-\varepsilon_LV$ in Eq.~(\ref{F1}) could be disregarded via RWA. Then, we simplify the system Hamiltonian in the squeezed frame as
\begin{equation}\label{F2}
\begin{split}
\hat{H}_{\mathrm{Total}}^{\mathrm{S}'}\simeq &\delta_{s} \hat{a}_{s}^{\dagger} \hat{a}_{s}+\delta_{ed}\hat{\sigma}_{+}\hat{\sigma}_{-} +g_{\mathrm{eff}}'(\hat{a}_{s}^{\dagger}\hat{\sigma}_{-}+\hat{a}_{s}\hat{\sigma}_{+})\\
&+\varepsilon_{L}'\left(\hat{a}_{s}^{\dagger}e^{-i\delta_L t}+\hat{a}_{s}e^{i\delta_L t}\right),
\end{split}
\end{equation}
with $g_{\mathrm{eff}}'=g_0\cosh(r_p)$ and $\varepsilon_L'=\varepsilon_L\cosh(r_p)$. Based on the effective Hamiltonian (\ref{F2}), PB can be achieved in case of appropriate parameters. Specifically, under the weak driving $\varepsilon_{L}'$, the system initially prepared in the ground state $|0,\mathcal{D}\rangle_{s}$ can be driven to the bare state $|1,\mathcal{D}\rangle_{s}$, which can then be dressed by the strong single-phonon spin-mechanical interaction ($g_{\mathrm{eff}}'\gg\varepsilon_{L}'$), yielding the first excited states of the system $|1,\pm\rangle_{s}\equiv\left( |1,\mathcal{D}\rangle_{s}\pm|0,\mathcal{E}\rangle_{s}\right) /\sqrt{2}$, with the energy splitting $2g_{\mathrm{eff}}'$. When $\delta_{s}-\delta_{L}=\pm g_{\mathrm{eff}}'$, the weak driving $\varepsilon_{L}'$ is resonantly coupled to the
transition $|0,\mathcal{D}\rangle_{s}\rightarrow|1,\pm\rangle_{s}$, while the transition $|1,\pm\rangle_{s}\rightarrow|2,\pm\rangle_{s}$
is detuned and thus suppressed for  $g_{\mathrm{eff}}'\gg\gamma_{m_\mathrm{{eff}}}$. This enables the appearance of single PB in the hybrid system with single-phonon spin-mechanical coupling. In addition, it is also possible to realize enhanced PB by tuning MPA that could induce amplified energy-spectrum anharmonicity.

\begin{figure}
	\includegraphics[width=1\columnwidth]{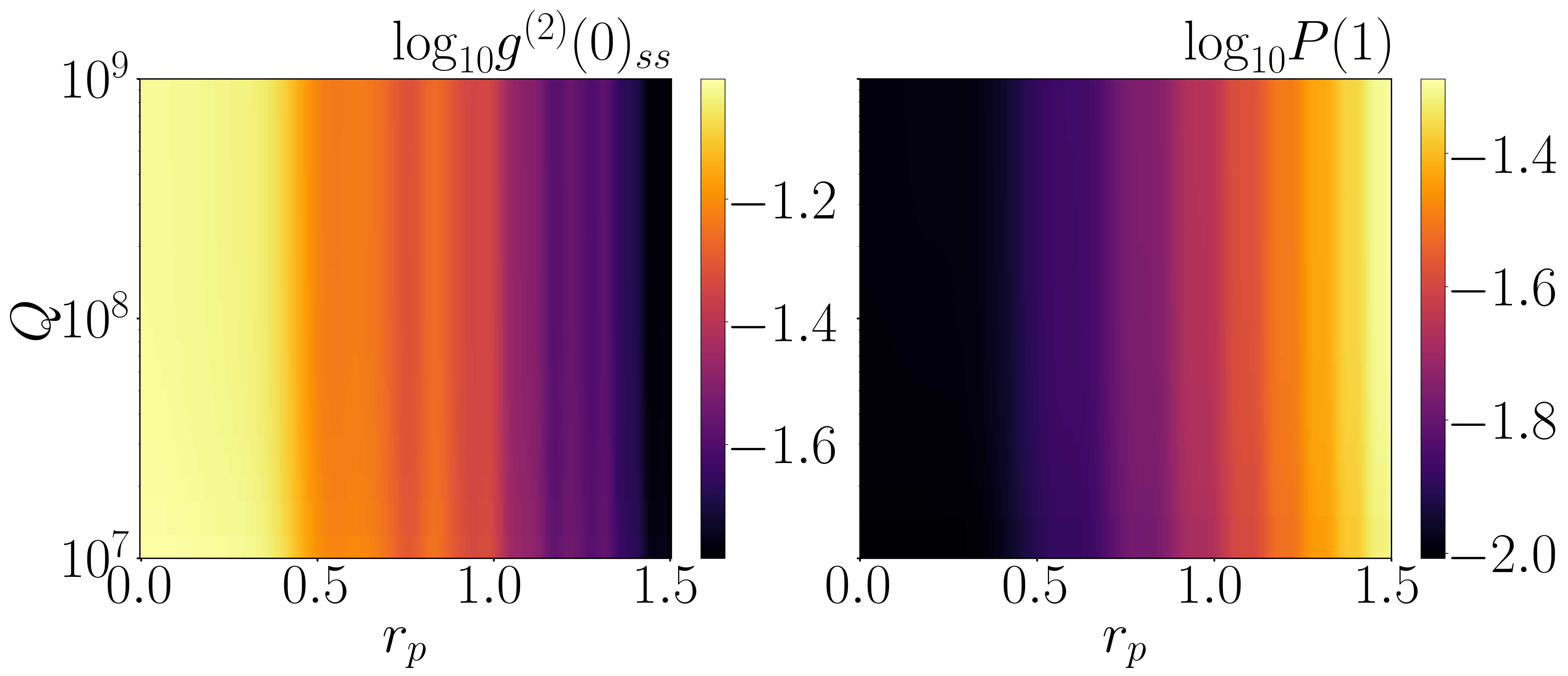}
	\caption{{  Logarithmic second-order correlation function $g^{(2)}(0)_{ss}$ and single-phonon probability $P(1)$ versus the mechanical oscillator's $Q$ factor and the squeezing parameter $r_p$. Parameters used here are $g_0=2\pi \times 2.5$  kHz, $\gamma_z=2\pi\times10$ Hz, $\varepsilon_L=0.01g_0$.}}
	\label{figA2}
\end{figure}

Here we choose $\delta_s=\delta_{ed}=50 g_{0}V$ for RWA, which ensures that the effective Hamiltonian (\ref{F2}) exactly reproduces the dynamics of the system described by the total Hamiltonian (\ref{F1}). In this case, the strength of MPA for achieving enhanced PB is limited to a relatively small range due to the restriction on system parameters imposed by the RWA condition. To be specific, we take a small squeezing parameter of $r_p=1.5$ as an example, which gives $\delta_s/2\pi\approx0.3$ MHz and thus $\delta_m/2\pi\approx3$ MHz, for $g_{0}/2\pi=2.5$ kHz. Such a value of $\delta_m$ is very close to the typical frequency of the mechanical oscillator we are considering. For stronger MPA with $r_p>1.5$, the value of $\delta_m$ becomes larger than the oscillator frequency, causing a contradiction with the parameter relation $\delta_{m}=\omega_{m}-\omega_{p}$. Given this, we limit the following discussions to a small-amplification regime with $r_p\le1.5$. Note that our scheme based on the two-phonon coupling can readily work in the large-amplification regime, at least for $r_p=3$, by virtue of the large difference between the bare two-phonon coupling rate and the oscillator frequency (about seven orders of magnitude), thus providing an attractive platform to access PB with better performance and large-scale tunability. 

Fig.~\ref{figA2} shows the logarithmic second-order correlation function $g^{(2)}(0)_{ss}$ and single-phonon probability $P(1)$ versus $r_p$ as well as the oscillator's $Q$ factor. Here the variation range of $Q$ is set to be consistent with that used in Fig.~\ref{fig6} for comparison. As expected, we observe that the performance of PB can be enhanced by increasing the strength of MPA. We obtain $g^{(2)}(0)_{ss}<10^{-1.6}$ and $P(1)>10^{-1.4}$ when $r_p\rightarrow1.5$. Further increasing the strength of the weak drive $\varepsilon_L$ could increase the single-phonon probability, while in turn the second-order correlation $g^{(2)}(0)_{ss}$ would be increased accordingly.
}

\begin{figure}[b]
	\includegraphics[width=1\columnwidth]{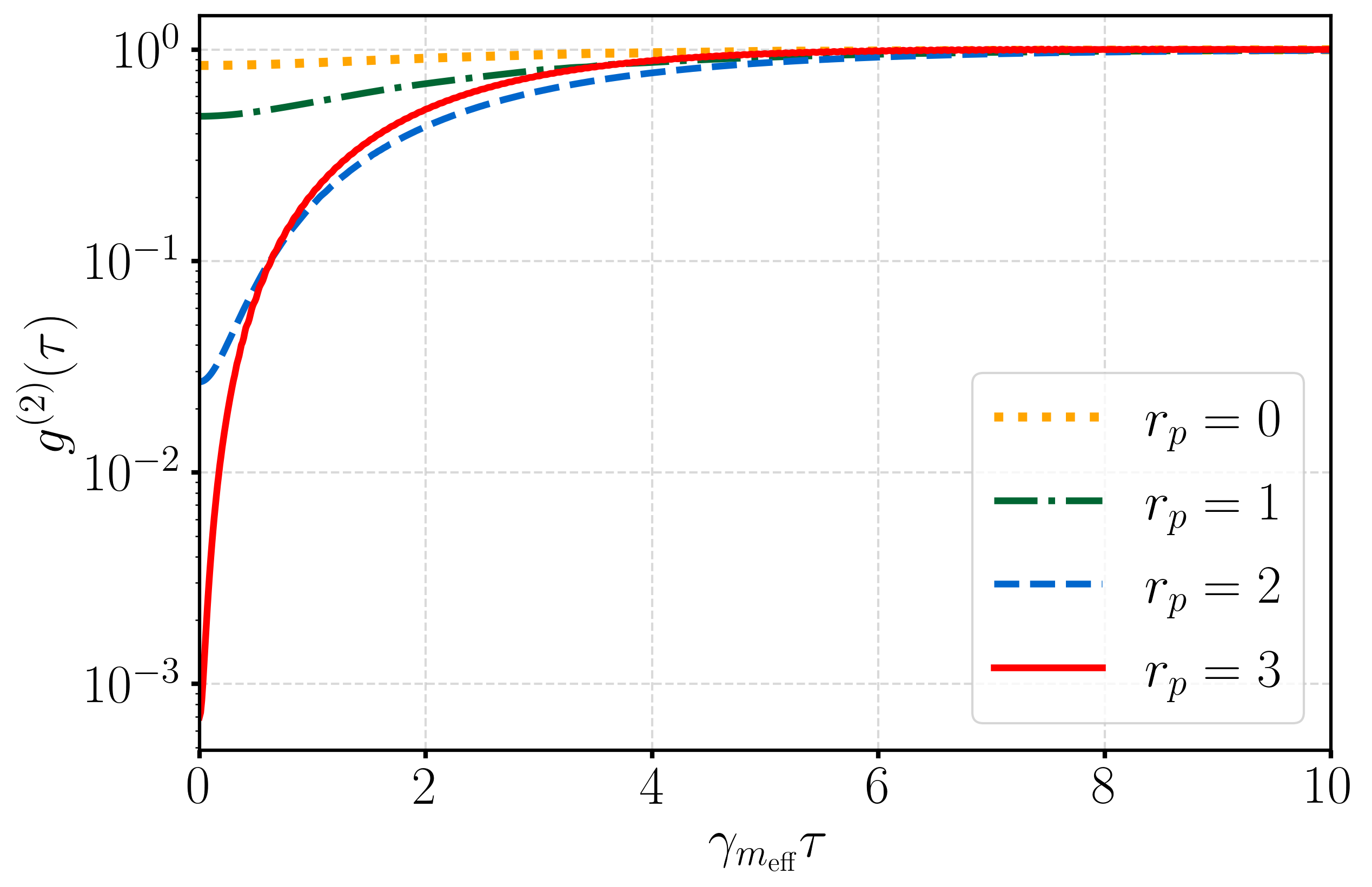}
	\caption{Steady-state delayed-time second-order correlation function $g^{(2)}(\tau)_{ss}$ versus the scaled time delay $\gamma_{m_\mathrm{{eff}}} \tau$ for different mechanical amplification parameter $r_p$. The parameters used here are the same as those for point A in Fig.~\ref{fig6}(c).}
	\label{figA3}
\end{figure}

\section{Phonon correlations with finite-time delays}\label{appenI}
As one of the quantum signatures for PB, we discuss briefly the phonon intensity correlations with finite-time delays. We calculate the delayed-time second-order correlation function $g^{(2)}(\tau)_{ss}$ according to the definition given in Eq.~(\ref{delayed g2}) and plot $g^{(2)}(\tau)_{ss}$ with respect to various mechanical amplification parameter $r_p$ in Fig.~\ref{figA3}. It shows that $g^{(2)}(\tau)_{ss}>g^{(2)}(0)_{ss}$ for arbitrary time delay $\tau$, manifesting the occurrence of phonon antibunching effect and the fact that phonon tends to be emitted singly rather than in groups. As the delay time increases gradually, the delayed-time second-order correlation function settles to 1 due to the loss of quantum coherence, and the phonon characterizes the standard Poisson distribution.

\bibliography{sample.bib}

\end{document}